\documentclass[pdflatex,bst/sn-mathphys-num,iicol]{sn-jnl}

\usepackage[T1]{fontenc}

\usepackage{graphicx}%
\usepackage{multirow}%
\usepackage{amsmath,amssymb,amsfonts}%
\usepackage{amsthm}%
\usepackage{mathrsfs}%
\usepackage[title]{appendix}%
\usepackage{xcolor}%
\usepackage{textcomp}%
\usepackage{manyfoot}%

\usepackage[natbib,style=numeric-comp,sorting=none]{biblatex}
\usepackage{booktabs}%
\usepackage{algorithm}%
\usepackage{algorithmicx}%
\usepackage{algpseudocode}%

\usepackage{soul}

\usepackage{multicol}
\usepackage{hyperref}
\usepackage{cleveref}
\usepackage{listings}

\usepackage{braket} 

\newcommand{\cudacpp}{\texttt{CUDACPP}}

\newcommand{\madgraph}{\textsc{MadGraph5\_aMC@NLO}}
\newcommand{\mg}{\textsc{MG5aMC}}

\newcommand{\rex}{\texttt{Rex}}
\newcommand{\tearex}{\texttt{teaRex}}

\newcommand{\madtrex}{\textsc{MadtRex}}

\newcommand{\codett}[1]{\lstinline[language=rex]!#1!}

\newcommand{\double}{\codett{double}}
\newcommand{\event}{\codett{event}}
\newcommand{\evbelongs}{\codett{eventBelongs}}
\newcommand{\particle}{\codett{particle}}
\newcommand{\parton}{\codett{parton}}
\newcommand{\process}{\codett{process}}
\newcommand{\lhe}{\codett{lhe}}
\newcommand{\weightor}{\codett{weightor}}
\newcommand{\iterator}{\codett{iterator}}
\newcommand{\procrwgt}{\codett{procReweightor}}
\newcommand{\rwgtr}{\codett{reweightor}}

\newcommand{\pointer}[1]{$^*$\codett{#1}}

\newcommand{\rwgtslha}{\codett{rwgt\_slha}}
\newcommand{\paramrwgt}{\codett{param\_rwgt}}

\newcommand{\oz}{\texttt{-O0}}
\newcommand{\oo}{\texttt{-O1}}
\newcommand{\otw}{\texttt{-O2}}
\newcommand{\oth}{\texttt{-O3}}

\newcommand{\m}{\mathcal{M}}
\newcommand{\msq}{\left| \mathcal{M}\right|^2}

\newcommand{\mprsq}{\left| \mathcal{M}' \right|^2 }

\lstdefinestyle{numbers} {numbers=left, stepnumber=1, numberstyle=\tiny, numbersep=10pt}

\lstdefinelanguage{madgraph}
{
    keywords=[1]{
    CALL, DO, END
    },
    keywordstyle=[1]\color[HTML]{228B22},
    keywords=[2]{
        VXXXXX, OXXXXX, IXXXXX, VVV1P0_1, FFV1_0
    },
    keywordstyle=[2]\color[HTML]{1027ad},
    basicstyle=\ttfamily\linespread{4},
    breaklines=false,
    columns=flexible,
    commentstyle=\color[rgb]{0.127,0.427,0.514}\ttfamily\itshape,
    escapechar=@,
    extendedchars=true,
    inputencoding=latin1,
    numbers=left,
    numberstyle=\tiny,
    comment=[l]{!},
    numbers=left,
    numberstyle=\tiny,
}

\lstdefinelanguage{cudacpp}
{
    keywords=[1]{
    M_ACCESS, W_ACCESS, CD_ACCESS, A_ACCESS 
    },
    keywordstyle=[1]\color[HTML]{228B22},
    keywords=[2]{
        vxxxxx, oxxxxx, ixxxxx, VVV1P0_1, FFV1_0
    },
    keywordstyle=[2]\color[HTML]{1027ad},
    basicstyle=\ttfamily\linespread{4},
    breaklines=false,
    columns=flexible,
    commentstyle=\color[rgb]{0.127,0.427,0.514}\ttfamily\itshape,
    escapechar=@,
    extendedchars=true,
    inputencoding=latin1,
    numbers=left,
    numberstyle=\tiny,
    comment=[l]{//},
    numbers=left,
    numberstyle=\tiny,
}

\lstdefinelanguage{terminal}
{
    alsoletter={._},
    keywords=[1]{
    cd, ln, git, generate, output, make, check_cpp.exe, launch, install
    },
    keywordstyle=[1]\color[HTML]{228B22},
    keywords=[2]{
        clone, s, BACKEND, FPTYPE, USEOPENMP, standalone_cudacpp, madevent_simd, madevent_gpu
    },
    keywordstyle=[2]\color[HTML]{1027ad},
    basicstyle=\ttfamily\linespread{4},
    breaklines=false,
    columns=flexible,
    commentstyle=\color[rgb]{0.127,0.427,0.514}\ttfamily\itshape,
    escapechar=@,
    extendedchars=true,
    inputencoding=latin1,
    comment=[l]{\#},
    literate={.git}{.git}4,
}

\lstdefinelanguage{rex}
{
    keywords=[1]{
    std, REX, tea, using, namespace, for, if, else, class, private, return, const, static, template, typename, assert, operator, this, public, explicit, type
    },
    keywordstyle=[1]\color[HTML]{228B22},
    keywords=[2]{
        double, int, string, shared_ptr, xmlNode, vector, void, bool, string_view, string, function, EventRange, InitTx, EventTx, optional, HeaderRaw, HeaderTx, arr2, arr3, arr4, arrN, event, eventComparatorConfig, eventSorter, eventBelongs, particle, event_view, size_t, lheReader, lheWriter, xmlReader, xmlWriter, xmlRaw, nullopt, weightor, procReweightor, reweightor, transform, iterator, threadPool, istream, ostream, iostream, fstream, ifstream, ofstream, process, lhe, EventRawT, EventRaw, Attr, event_hash_fn, lheRaw, initNode, any, initTx, lheRaw, InitRaw, event_equal_fn, cevent_equal_fn, event_bool_fn, decay, decay_t, unsigned, short, long, param_rwgt, my_pdfs, parton, unordered_map, slha, xmlDoc, unique_ptr, HeaderRawOpt, array, mdspan, arrNRef, vecArrN, float, nStrideIter, complex, cArrN, cArrNRef, vecCArrN, mutex, cevent_hash_fn, thread, queue, rwgt_slha, event_comp_fn, EvRawT, EventRawT, char
    },
    keywordstyle=[2]\color[HTML]{1027ad},
    keywords=[3]{
        set, get, parse, attrs, add_attr, set_attr, emplace_back, read_lhe, write_lhe, set_event_comparator, load_lhef, transpose, print, set_status_filter, make_const_comparator, set_n, set_pdg, set_status, set_mass, set_event_comparator, belongs, set_indices, set_init_translator, xml_reader, xml_to_init, xml_to_any, xml_writer, init_to_xml, event_to_xml, header_to_xml, to_xml_raw, xml_writer, to_raw, set_event_translator, set_filter, move, is_same_v, read, evaluate, vec_elem_multi, append_wgts, set_event_checker, set_normaliser, set_reweight_functions, set_process, set_reweightors, add_reweightor, set_initialise, set_finalise, set_iterators, add_iterator, reserve, emplace_back, begin_batch, enqueue, foo, make_shared, push_back, write_rwgt_card, add_param, write, read_slha_rwgt, create, move_param_card, remove_param_card, get_card_writers, increment, qq_pdf, get_iterators, run, make_comparator, get_comp, size, main, make_reweightor, nUP, n, idPrUP, idPr, xWgtUP, xWgt, weight, scalUP, scale, aQEDUP, alphaQED, aQED, aEW, alphaEW, aQCDUP, aQCD, alphaS, aS, idUP, id, pdg, iStUP, iSt, status, mother, iColUP, iCol, icol, pUP, momentum, p, momenta, mUP, m, mass, vTimUP, vTim, vtim, spinUP, spin, mothUP, moth, flat_vector, at, E, px, py, pz, view, begin, end, add_particle, pT, eT, m, pL, pT2, eT2, mT2, pL2, phi, theta, rap, eta,  muF, muR, muPS, gS, events_to_processes, processes_to_events, idBmUP, idBm, eBmUP, eBm, pdfGUP, pdfG, pdfSUP, pdfS, idWgtUP, idWgt, nProcUP, nProc, xSecUP, xSec, xSecErrUP, xSecErr, xMaxUP, xMax, lProcUP, lProc, write_lhef, to_slha, load_slha, name, content, full, value, children, add_child, has_child, get_child, get_children, remove_child, replace_child, any_cast, has, set_tolerance, get_event_bool, position, set_header_translator, dot, set_hash, sort, initialise, evaluate, iterate, wait_batch, hardware_concurrency, card_writers, amp, external_legs_comparator, bridgeConstr, mT, set_x_translator
    },
    keywordstyle=[3]\color[HTML]{755E00},
    keywords=[4]{
    auto, true, false, constexpr, decltype, declval, cout, nullptr, nullptr_t, monostate, runtime_error, npos, get_hash, set_sorter
    },
    keywordstyle=[4]\color[HTML]{8a0101},
    basicstyle=\ttfamily\linespread{4},
    breaklines=false,
    columns=flexible,
    commentstyle=\color[rgb]{0.127,0.427,0.514}\ttfamily\itshape,
    escapechar=@,
    extendedchars=true,
    inputencoding=latin1,
    numbers=left,
    numberstyle=\tiny,
    comment=[l]{//},
    numbers=left,
    numberstyle=\tiny,
    stringstyle=\color[HTML]{00FFF2}\ttfamily,
    showstringspaces=false
}

\lstdefinelanguage{rexaster}
{
    keywords=[1]{
    std, REX, tea, using, namespace, for, if, else, class, private, return, const, static, template, typename, assert, operator, this, public, explicit, type
    },
    keywordstyle=[1]\color[HTML]{228B22},
    keywords=[2]{
        double, int, string, shared_ptr, xmlNode, vector, void, bool, string_view, string, function, EventRange, InitTx, EventTx, optional, HeaderRaw, HeaderTx, arr2, arr3, arr4, arrN, event, eventComparatorConfig, eventSorter, eventBelongs, particle, event_view, size_t, lheReader, lheWriter, xmlReader, xmlWriter, xmlRaw, nullopt, weightor, procReweightor, reweightor, transform, iterator, threadPool, istream, ostream, iostream, fstream, ifstream, ofstream, process, lhe, EventRawT, EventRaw, Attr, event_hash_fn, lheRaw, initNode, any, initTx, lheRaw, InitRaw, event_equal_fn, cevent_equal_fn, event_bool_fn, decay, decay_t, unsigned, short, long, param_rwgt, my_pdfs, parton, unordered_map, slha, xmlDoc, unique_ptr, HeaderRawOpt, array, mdspan, arrNRef, vecArrN, float, nStrideIter, complex, cArrN, cArrNRef, vecCArrN, mutex, cevent_hash_fn, thread, queue, rwgt_slha, event_comp_fn, EvRawT, EventRawT, char
    },
    keywordstyle=[2]\color[HTML]{1027ad},
    keywords=[3]{
        set, get, parse, attrs, add_attr, set_attr, emplace_back, read_lhe, write_lhe, set_event_comparator, load_lhef, transpose, print, set_status_filter, make_const_comparator, set_n, set_pdg, set_status, set_mass, set_event_comparator, belongs, set_indices, set_init_translator, xml_reader, xml_to_init, xml_to_any, xml_writer, init_to_xml, event_to_xml, header_to_xml, to_xml_raw, xml_writer, to_raw, set_event_translator, set_filter, move, is_same_v, read, evaluate, vec_elem_multi, append_wgts, set_event_checker, set_normaliser, set_reweight_functions, set_process, set_reweightors, add_reweightor, set_initialise, set_finalise, set_iterators, add_iterator, reserve, emplace_back, begin_batch, enqueue, foo, make_shared, push_back, write_rwgt_card, add_param, write, read_slha_rwgt, create, move_param_card, remove_param_card, get_card_writers, increment, qq_pdf, get_iterators, run, make_comparator, get_comp, size, main, make_reweightor, nUP, idPrUP, idPr, xWgtUP, xWgt, weight, scalUP, scale, aQEDUP, alphaQED, aQED, aEW, alphaEW, aQCDUP, aQCD, alphaS, aS, idUP, id, pdg, iStUP, iSt, status, mother, iColUP, iCol, icol, pUP, momentum, p, momenta, mUP, m, mass, vTimUP, vTim, vtim, spinUP, spin, mothUP, moth, flat_vector, at, E, px, py, pz, view, begin, end, add_particle, pT, eT, m, pL, pT2, eT2, mT2, pL2, phi, theta, rap, eta,  muF, muR, muPS, gS, events_to_processes, processes_to_events, idBmUP, idBm, eBmUP, eBm, pdfGUP, pdfG, pdfSUP, pdfS, idWgtUP, idWgt, nProcUP, nProc, xSecUP, xSec, xSecErrUP, xSecErr, xMaxUP, xMax, lProcUP, lProc, write_lhef, to_slha, load_slha, name, content, full, value, children, add_child, has_child, get_child, get_children, remove_child, replace_child, any_cast, has, set_tolerance, get_event_bool, position, set_header_translator, dot, set_hash, sort, initialise, evaluate, iterate, wait_batch, hardware_concurrency, card_writers, amp, external_legs_comparator, bridgeConstr, mT, set_x_translator
    },
    keywordstyle=[3]\color[HTML]{755E00},
    keywords=[4]{
    auto, true, false, constexpr, decltype, declval, cout, nullptr, nullptr_t, monostate, runtime_error, npos, get_hash, set_sorter
    },
    keywordstyle=[4]\color[HTML]{8a0101},
    basicstyle=\ttfamily\linespread{4},
    breaklines=false,
    columns=flexible,
    commentstyle=\color[rgb]{0.127,0.427,0.514}\ttfamily\itshape,
    escapechar=@,
    extendedchars=true,
    inputencoding=latin1,
    numbers=left,
    numberstyle=\tiny,
    comment=[l]{//},
    numbers=left,
    numberstyle=\tiny,
    stringstyle=\color[HTML]{00FFF2}\ttfamily,
    showstringspaces=false
}

\lstdefinelanguage{rexstar}
{
    keywords=[1]{
    std, REX, tea, using, namespace, for, if, else, class, private, return, const, static, template, typename, assert, operator, this, public, explicit, type
    },
    keywordstyle=[1]\color[HTML]{228B22},
    keywords=[2]{
        double, int, string, shared_ptr, xmlNode, vector, void, bool, string_view, string, function, EventRange, InitTx, EventTx, optional, HeaderRaw, HeaderTx, arr2, arr3, arr4, arrN, event, eventComparatorConfig, eventSorter, eventBelongs, particle, event_view, size_t, lheReader, lheWriter, xmlReader, xmlWriter, xmlRaw, nullopt, weightor, procReweightor, reweightor, transform, iterator, threadPool, istream, ostream, iostream, fstream, ifstream, ofstream, process, lhe, EventRawT, EventRaw, Attr, event_hash_fn, lheRaw, initNode, any, initTx, lheRaw, InitRaw, event_equal_fn, cevent_equal_fn, event_bool_fn, decay, decay_t, unsigned, short, long, param_rwgt, my_pdfs, parton, unordered_map, slha, xmlDoc, unique_ptr, HeaderRawOpt, array, mdspan, arrNRef, vecArrN, float, nStrideIter, complex, cArrN, cArrNRef, vecCArrN, mutex, cevent_hash_fn, thread, queue, rwgt_slha
    },
    keywordstyle=[2]\color[HTML]{1027ad},
    keywords=[3]{
        set, get, parse, attrs, add_attr, set_attr, emplace_back, read_lhe, write_lhe, set_event_comparator, load_lhef, transpose, print, set_status_filter, make_const_comparator, set_n, set_pdg, set_status, set_mass, set_event_comparator, belongs, set_indices, set_init_translator, xml_reader, xml_to_init, xml_to_any, xml_writer, init_to_xml, event_to_xml, header_to_xml, to_xml_raw, xml_writer, to_raw, set_event_translator, set_filter, move, is_same_v, read, evaluate, vec_elem_multi, append_wgts, set_event_checker, set_normaliser, set_reweight_functions, set_process, set_reweightors, add_reweightor, set_initialise, set_finalise, set_iterators, add_iterator, reserve, emplace_back, begin_batch, enqueue, foo, make_shared, push_back, write_rwgt_card, add_param, write, read_slha_rwgt, create, move_param_card, remove_param_card, get_card_writers, increment, qq_pdf, get_iterators, run, make_comparator, get_comp, size, main, make_reweightor, nUP, n, idPrUP, idPr, xWgtUP, xWgt, weight, scalUP, scale, aQEDUP, alphaQED, aQED, aEW, alphaEW, aQCDUP, aQCD, alphaS, aS, idUP, id, pdg, iStUP, iSt, status, mother, iColUP, iCol, icol, pUP, momentum, p, momenta, mUP, m, mass, vTimUP, vTim, vtim, spinUP, spin, mothUP, moth, flat_vector, at, E, px, py, pz, view, begin, end, add_particle, pT, eT, m, pL, pT2, eT2, mT2, pL2, phi, theta, rap, eta,  muF, muR, muPS, gS, events_to_processes, processes_to_events, idBmUP, idBm, eBmUP, eBm, pdfGUP, pdfG, pdfSUP, pdfS, idWgtUP, idWgt, nProcUP, nProc, xSecUP, xSec, xSecErrUP, xSecErr, xMaxUP, xMax, lProcUP, lProc, write_lhef, to_slha, load_slha, name, content, full, value, children, add_child, has_child, get_child, get_children, remove_child, replace_child, any_cast, has, set_tolerance, get_event_bool, position, set_header_translator, dot, set_hash, sort, evaluate, iterate, wait_batch, hardware_concurrency, card_writers, set_x_translator
    },
    keywordstyle=[3]\color[HTML]{755E00},
    keywords=[4]{
    auto, true, false, constexpr, decltype, declval, cout, nullptr, nullptr_t, monostate, runtime_error, npos, get_hash, set_sorter
    },
    keywordstyle=[4]\color[HTML]{8a0101},
    basicstyle=\ttfamily\linespread{4},
    breaklines=false,
    columns=flexible,
    commentstyle=\color[rgb]{0.127,0.427,0.514}\ttfamily\itshape,
    escapechar=@,
    extendedchars=true,
    inputencoding=latin1,
    numbers=left,
    numberstyle=\tiny,
    comment=[l]{//},
    numbers=left,
    numberstyle=\tiny,
    stringstyle=\color[HTML]{00FFF2}\ttfamily,
    showstringspaces=false
}

\theoremstyle{thmstyleone}%
\theoremstyle{thmstyletwo}%

\theoremstyle{thmstylethree}%

\begin{document}

\title[Rapid event extraction and tensorial event adaption]{Rapid event extraction and tensorial event adaption}
\subtitle{Libraries for efficient access and generic reweighting of parton-level events and their implementation in the \madtrex{} module}

\author[1]{\fnm{Stefan} \sur{Roiser}}\email{stefan.roiser@cern.ch}

\author[2]{\fnm{Robert} \sur{Sch{\"o}fbeck}}\email{robert.schoefbeck@oeaw.ac.at}

\author*[1,2]{\fnm{Zenny} \sur{Wettersten}}\email{zenny.wettersten@cern.ch}

\affil[1]{\orgname{CERN}, \orgaddress{\street{Esplanade des Particules 1}, \city{Geneva}, \postcode{1211}, \country{Switzerland}}}

\affil[2]{\orgdiv{MBI (HEPHY)}, \orgname{{\"O}AW}, \orgaddress{\street{Dominikanerbastei 16}, \city{Vienna}, \postcode{1010}, \country{Austria}}}


\abstract{We present \rex{} and \tearex{}, C++17 libraries for efficient management of parton-level hard scattering event information and completely generic reweighting of such events, respectively. \rex{} is primarily an interfacing and I/O library for Les Houches Event format files and provides an internal event format designed with data parallelism in mind, and \tearex{} extends this format to provide full parton-level reweighting functionality with minimal code needing to be written by the end user. These libraries serve as the foundation for the \madtrex{} reweighting module for \madgraph{}, extending the functionality of the \cudacpp{} plugin to allow for data-parallel model-generic leading order parameter reweighting on SIMD-enabled CPUs and SIMT GPUs, speeding up reweighting by more than two orders of magnitude compared to \madgraph{} running on the exact same hardware while providing trivial scalability to larger and distributed systems.}




\maketitle

\section{Introduction}
\label{sec:intro}

Adoption of explicit data-parallel hardware acceleration for high-energy physics (HEP) software --- using on-CPU SIMD instructions and off-CPU SIMT GPU offloading --- has in recent years not only proven to be of great importance in the face of impending computational needs, but also a very difficult task \cite{HEPSoftwareFoundation:2017ggl,CERN-LHCC-2022-005,Software:2815292,HSFPhysicsEventGeneratorWG:2020gxw}. Significant work has been put into porting existing or writing new HEP software to properly utilise existing and upcoming hardware \cite{Kanzaki:2010ym,Hagiwara:2010oca,Hagiwara:2013oka,Bothmann:2021nch,Carrazza:2021gpx,Bothmann:2023gew,Cruz-Martinez:2025kwa,Valassi:2021ljk,Valassi:2022dkc,Valassi:2023yud,Hageboeck:2023blb,Valassi:2025xfn}. However, these efforts generally target specific codebases and implementations, leaving common issues such as the interfacing between older, typically object-oriented (OO) data formats and structures-of-arrays (SoA), more fit for data parallelism, a problem re-implemented across different collaborations.

Simultaneously, improved experimental precision and the resulting increase in necessary theoretical event samples for statistical comparisons with the standard model (SM) make the efficient reuse of event samples ever more important. In SM studies this can occur e.g. when estimating simulation uncertainties by evaluating different parton distribution function (pdf) sets for the initial-state particles in a parton-level hard scattering event and varying the factorisation and renormalisation scales they are evaluated at; alternatively, when studying the phenomenology of beyond-the-SM (BSM) models with their infinite available parameter spaces it is unfeasible to run the full simulation chain for samples at all parameters of all models, and instead the Monte-Carlo event weights of an existing sample can be re-evaluated for the new model and propagated to the end experimental observables in a procedure known as matrix element reweighting (henceforth called parameter reweighting).

We attempt to address both of these issues with two new C++17 libraries: \rex{}, the \texttt{R}apid \texttt{e}vent e\texttt{x}traction library; and \tearex{}, a library for \texttt{t}ensorial \texttt{e}vent \texttt{a}daption with \rex{}. The former is intended to provide interfacing between OO data formats following the conventions of the Les Houches Event (LHE) file format and SoA formats with event data stored in contiguous vectors, in \rex{} split into independent SoAs for individual subprocesses based on user-provided event categorisation. On the other hand, \tearex{} is a minimal extension to \rex{} providing a structure for generic parton-level event reweighting with minimal interfacing necessary from developers, using the event sorting capabilities from \rex{} alongside its SoA event structure to enable immediate SIMD- and SIMT-friendly data access to event data to automate the full reweighting process in a completely generic way for any user-supplied reweighting function whether it be for parameter, pdf, or any other type of reweighting. Additionally, \rex{} and \tearex{} are used as a foundation for the \madtrex{} module, which repurposes the data-parallel scattering amplitudes generated by the \cudacpp{} plugin \cite{Valassi:2021ljk,Valassi:2022dkc,Valassi:2023yud,Hageboeck:2023blb,Valassi:2025xfn,Hagebock:2025jyk} for \madgraph{} (\mg{}) \cite{Alwall:2014hca} to create executables for generic tree-level parameter reweighting. Using on-CPU SIMD instructions and GPU offloading, \madtrex{} increases peak reweighting throughput for computationally heavy processes by more than two orders of magnitude when compared to \mg{}, but even without any explicitly implemented data parallelism on-CPU \madtrex{} executables without SIMD instructions increase event throughput by roughly a factor $30-50$ due to a combination of the better-scaling sorting algorithm used by \rex{} and \tearex{}, running through a compiled executable rather than an interpreted language, and compiler optimisations including but not limited to automatically applied multithreading.

This paper is split into three main sections: \cref{sec:rex} provides a detailed description of the \rex{} library, a usage manual for applying it to other C++ programs, and benchmarks for the included LHE file format reader and sorting algorithm; \tearex{} is then presented in \cref{sec:tearex}, also with a usage manual as well as a sketch for how to apply it specifically for pdf reweighting; and finally a usage guide for \madtrex{} is given in \cref{sec:madtrex} with throughput comparisons to the default \mg{} reweighting module as applied to BSM reweighting specifically in the SM Effective Field Theory (SMEFT) using a SMEFTsim UFO model \cite{Brivio:2017btx,Brivio:2020onw}. We finish the paper with a summary alongside discussion regarding possible future development directions and considerations for all three presented codes in \cref{sec:conclusions}.

\section{Rapid event extraction}
\label{sec:rex}

The Les Houches Events file format (LHE) \cite{Alwall:2006yp,Butterworth:2010ym,Andersen:1699963} is a human-readable XML-based format for storing parton-level hard scattering event information intended for interfacing between high energy physics (HEP) software in a generic yet simple way. However, despite the shared input/output (I/O) format, parton-level event generators have generally designed their own interfaces for matching their internal data formats to the read or written LHE file. The \texttt{Rapid event extraction} C++ library (\rex{}) is intended to serve as a simple and efficient tool for reading and writing LHE files, while providing a data-oriented internal data format with memory laid out specifically to simplify the matching between the natively object-oriented LHE format and modern event-level data-parallel programs.

To facilitate this goal, \rex{} has two internal storage formats which can easily be transposed between with a single function call: An XML-adjacent tree structure where each event is stored as a separate \event{} object, and a structure-of-arrays (SoA) format of \process{} objects where event data are merged into singular, contiguous arrays based on user-input sorting functions. While this latter structure was designed for the event-level data-parallel format of the \cudacpp{} plugin \cite{Valassi:2021ljk,Valassi:2022dkc,Valassi:2023yud,Hageboeck:2023blb,Valassi:2025xfn,Hagebock:2025jyk} for \madgraph{} \cite{Alwall:2014hca}, we expect the functionality to be applicable to any current or future software looking to implement a data-parallel multi-event interface due to the necessity of data-oriented formats for proper utilisation of data-parallel hardware such as SIMD-enabled CPUs and SIMT GPU\footnote{We forego details on SIMD and SIMT parallelism as well as the differences between OO arrays of structures and SoAs here, but a quick internet search will provide extensive explanations. As a short description, arrays of structures can be thought of as objects used in chronologically ordered for-loops, while SoAs are flipped such that equivalent data are adjacent in memory to allow the same for-loop to be ordered in space rather than time; with this in mind, SIMD and SIMT architectures allow for this exact type of data parallelism where the same compute instruction is performed across many equivalent data at the same time.}.

Furthermore, \rex{} was designed with modularity in mind, both with respect to what data is needed within a given software and with respect to what file format the underlying event data is stored in on disk. As of version 1.0.0 \rex{} only has native support for the XML-based LHE v3.0 format, but the internal data format and the I/O routines are completely disparate yet simple to interface, making extensions to other formats such as the HDF5-based \cite{The_HDF_Group_Hierarchical_Data_Format} LHEH5 format \cite{Hoche:2019flt,Bothmann:2023ozs} minimal and possible by users with minimal necessary interfacing --- the \codett{lheReader} class can be constructed from a function constructing \event{} objects as well as one constructing \codett{initNode} objects, which hold the process information corresponding to the \codett{<init>} node in the LHE standard. The \codett{lheWriter}, mapping \lhe{} objects to a user-provided format, can be implemented similarly.

\rex{} was designed with the principal goal of simplifying efficient HEP software design while maintaining physics-driven data access with a generic base structure allowing both for the immediate use of \rex{} features and an extensive, adaptable interface for advanced use cases, all the while providing sufficient internal support for any level of complexity in between these extremes. The functionality of \rex{} can roughly be grouped into three categories based on these principles:
\begin{itemize}
    \item \textbf{Physics-oriented data access:} LHE-based objects, including the \event{}, \process{}, and \lhe{} structs, intended to give immediate access to physics data according to the LHE standard, with zero interfacing necessary other than that necessary to load an LHE file.  These objects provide immediate access to all the underlying data by reference, such that they can be directly modified without needing to create and set data with additional function calls.
    \item \textbf{Helper classes for customisation:} Relatively simple wrappers for e.g. constructing event comparison functions, storing and accessing additional event-level information not part of the LHE standard, such as pdf information, or translation to and from other data formats. While they do not provide full customisability, these wrappers make it easy to set up more specific configurations than just immediate object-oriented event access. 
    \item \textbf{Bare data and functionality:} Underlying fundamental data types and the templated base classes used to define them, as well as the non-wrapped function types the helper wrappers discussed above give access to, such as completely generic event sorters.
\end{itemize}

In the manual provided in \cref{sec:rex_manual}, these are presented in the order listed above, starting with the plug-and-play access to LHE format data in a C++ program without consideration for underlying data handling and ending with descriptions of the underlying data types and the definitions of interfacing functionality. Then, some simplified illustrative implementations and use cases are shown in \cref{sec:rex_usecase} in the same order.

\subsection{Manual}
\label{sec:rex_manual}

This section is intended to provide a practical user manual for the \rex{} library, describing its structure and usage. Due to its intentional simple-to-complex and specific-to-generic design, this manual is split into three separate parts: the first, \cref{sec:rex_dataaccess}, describes the default data access format for interfacing with LHE-style data using the \event{}, \process{}, and \lhe{} types as well as loading and writing LHE standard files; the second section provides an introduction to the functionality wrappers allowing for customisation without the need for defining comparators, sorters, and type translators from scratch; finally, \cref{sec:rex_fundamentals} gives a description of the underlying data types and storage formats to allow power users full generic applicability of \rex{} with completely generic transposition between OO- and SoA-formats. For users just looking to use \rex{} to read and write the XML-based LHE format and simplify existing workflows, we recommend reading the first two parts, while power users looking to integrate a standardised data format for parton-level HEP information may also find interest in \cref{sec:rex_fundamentals}.

\subsubsection{Physics-driven data access}
\label{sec:rex_dataaccess}

\begin{table}[t]
    \centering
    \begin{tabular}{c|c|c}
        \textbf{Data}& \textbf{Type} & \textbf{Access functions}\\ \hline
         No. partons & \codett{size\_t} & \codett{nUP()}, \codett{n()} \\
         Process index & \codett{long int} & \codett{idPrUP()}, \codett{idPr()} \\
         Event weight & \double{} & \codett{xWgtUP()}, \codett{xWgt()}, \codett{weight()}\\
         Event scale & \double{} & \codett{scalUP()}, \codett{scale()}\\ 
         $\alpha_{EW}$ & \double{} & \begin{tabular}{c}
              \codett{aQEDUP()}, \codett{alphaQED()},\\ \codett{aQED()}, \codett{aEW()}, \codett{alphaEW()}
         \end{tabular}\\
         $\alpha_{S}$ & \double{} & \begin{tabular}{c}
              \codett{aQCDUP()}, \codett{aQCD},\\ 
              \codett{alphaS()}, \codett{aS()}
         \end{tabular}
    \end{tabular}
    \caption{Event characterisation data as accessible through the \codett{event} type in \rex{}. All data are given by reference when accessed through the listed functions, meaning they can be modified directly through the access functions.}
    \label{tab:rex_event_data}
\end{table}

At its core, \rex{} is a library for accessing information according to the LHE standard while providing transposition between the OO-format given by the XML-based standard and a data-oriented SoA format where all event data is stored contiguously in memory, allowing for simple access to data parallelism using SIMD instructions and SIMT machines. The three types relevant for this purpose are the OO \event{} struct, the SoA \process{} struct, and the overarching \lhe{} struct.

From an access perspective, the \event{} type is quite simple --- it contains all event-level data part of the LHE standard. Although internally these are stored with different names than typical, the \event{} type has access function for many standard names for these variables, as detailed in \cref{tab:rex_event_data,tab:rex_parton_data} (internally, \rex{} uses custom types for arrays and vectors of arrays that ensure safety and contiguous storage --- elaborated on in \cref{sec:rex_fundamentals} --- but the important point to note is that e.g. vectors of four-momenta are stored contiguously and are accessed through a double index, the first referring to the line and the second to the momentum component; furthermore, \codett{std::vector}s of the corresponding information can be accessed through the member function \codett{flat\_vector()}). Additionally, note that individual partons (given as \particle{} objects) of an \event{} object can be accessed through indexing access operators \codett{operator[](...)} and \codett{at(...)}. The \particle{} type is a member of the \event{} type, which gives a view of that particular index of the data stored in the \event{} object, meaning \event{}s can be treated either as collections of parton-level data or as collections of partons with individual data. Partons additionally have reference access to their individual momentum components through the functions \codett{E(), px(), py(),} and \codett{pz()}. When treating four-momenta, note that \rex{} stores these in the $(t,x,y,z)$ basis with mass $m$ stored separately, as opposed to the LHE format where they are treated as five-dimensional arrays in the basis $(x,y,z,t,m)$.

\begin{table}[t]
    \centering
    \begin{tabular}{c|c|c}
       \textbf{Data}& \textbf{Type} & \textbf{Access functions}\\ \hline
        PDG code & \codett{long int}&  \codett{idUP()}, \codett{id()}, \codett{pdg()} \\
        Status & \codett{short int}  & \begin{tabular}{c}
              \codett{iStUP()}, \codett{iSt()},\\ \codett{status()}
        \end{tabular}  \\
        Mothers & \codett{short int[2]}  & \begin{tabular}{c}
              \codett{mothUP()}, \codett{moth()},\\ \codett{mother()} 
        \end{tabular}    \\
        Colour flow &  \codett{short int[2]}  & \begin{tabular}{c}
             \codett{iColUP()}, \codett{iCol()},\\ \codett{icol()}
        \end{tabular} \\
        Momentum$^*$  & \codett{double[4]}  &  \begin{tabular}{c}
             \codett{pUP()}, \codett{momentum()},\\
             \codett{p()}, \codett{momenta()} 
        \end{tabular}   \\
        Mass$^*$   & \double{}   & \codett{mUP()}, \codett{m()}, \codett{mass()} \\
        Lifetime & \double{}  &  \begin{tabular}{c}
             \codett{vTimUP()}, \codett{vTim()},\\ \codett{vtim()}
        \end{tabular}  \\
        Spin &  \double{}  &  \codett{spinUP()}, \codett{spin()}   
    \end{tabular}
    \caption{Parton-level data as accessible through the \event{} type in \rex{}. All data are given by reference when accessed through the listed functions, meaning they can be modified directly through the access functions. Additionally, views of individual parton lines are accessible through the \codett{particle} member of the \event{} type, which can also be accessed by the index operators \codett{operator[](...)} and \codett{at(...)} of \event{} objects. The \particle{} subtype has the exact same access functions as the \event{} type, although accessing it through the former provides the data for a single event parton while the latter provides a vector of all values for each particle in the event.\\ *Note that while the LHE format stores particle momenta as arrays of 5 doubles ordered according to the $(x,y,z,t,m)$ basis with parton mass $m$ appended as a fifth and final entry, \rex{} stores momenta in the $(t,x,y,z)$ basis and masses separately from the momenta.}
    \label{tab:rex_parton_data}
\end{table}

In addition to the direct data access provided by the functions mentioned above, \event{}s can be provided arbitrary parton orderings through the member \codett{std::vector<size\_t> indices} storing a given parton ordering, and the \codett{event\_view} member type of \event{}s accessible from the \event{} member function \codett{view()}, which overrides the indexing operator \codett{operator[]}  to index according to the \codett{indices} vector (possibly ignoring partons stored in the owning \event{}, depending on the ordering).

Additionally, individual particles can be stored in the \parton{} struct, which is an owning but otherwise equivalent type to the \particle{} struct. These can be used to create \event{} objects or to append partons to existing \event{}s using the \codett{event(std::vector<parton>)} constructor or \codett{add\_particle(const parton\&)} member functions, respectively. Both \parton{} and \particle{}s have additional member functions to calculate observables, such as transverse momentum \codett{pT()}, transverse energy \codett{eT()}, transverse mass \codett{mT()}, or longitudinal momentum \codett{pL()} (each of which also have a corresponding squared operator \codett{pT2()}, \codett{eT2()}, \codett{mT2()}, and \codett{pL2()}), or azimuthal angle \codett{phi()}, polar angle \codett{theta()}, rapidity \codett{rap()}, and pseudorapidity \codett{eta()}.

The \event{} type comes with self-returning setters for each non-derived variable mentioned above, given by member functions \codett{set\_...}. These do not come overloaded with different names (although this could be implemented should it be desired), so exact names should be read from the header \codett{Rex.h}. Events also have a vector \codett{wgts\_} which stores any additional event weights --- relevant for \tearex{} reweighting --- as well as a shared vector of weight labels \codett{weight\_ids} which is primarily meant to be accessed from an owning \lhe{} object. Finally, specific scales $\mu_{F,R,PS}$ for refactorisation, renormalisation, and parton showers, are provided, although they are not mandated by the LHE standard, and can either be accessed directly through the corresponding functions \codett{muF()}, \codett{muR()}, or \codett{muPS()} (although these functions directly return a reference to the corresponding double which may be equal to 0), but can also be accessed through \codett{get\_...} functions which first check whether the particular value is zero and return \texttt{scalUP} if it is.

Collections of events are transposed into the \process{} type, which has all the corresponding members as an \event{} object with the difference that every variable is stored in a contiguous vector, not just parton-level data. This includes the scales $\mu$ mentioned above. All these vectors can be accessed through identically named member functions as for \event{}s. \process{} objects additionally have access to a vector of shared pointers to \event{}s, intended to be defined by \lhe{} objects at transposition, and include (self-returning) transposition functions \codett{transpose\_...} for all contiguous vector quantities to map these quantities back into corresponding events without needing to transpose all data. These transposition functions are overloaded for all naming schemes shown in \cref{tab:rex_event_data,tab:rex_parton_data}, as well as a total transposition operator \codett{transpose()} which resets the vector of owned \event{}s (without necessarily deleting existing \event{}s, depending on scope) with new ones defined from the owned contiguous vectors. It is worth mentioning also here that the array-like types used to store \texttt{momenta}, \texttt{iCol}, and \texttt{mother} can be accessed as \codett{std::vector}s of the corresponding type using the member function \codett{flat\_vector()}, which returns reference access to a vector of the data now lacking the double indices.

Both \event{} and \process{} objects have an additional \codett{extra} member, which is an unordered map from labels given as \codett{std::string} to the given value stored as \codett{std::any} for \event{}s and \codett{std::vector<std::any>} for \process{}es. These are detailed further in \cref{sec:rex_wrappers}. Finally, $g_S$ can be calculated directly from $\alpha_S$ using member functions \codett{gS()} which return a \double{} for \event{}s and \codett{std::vector<double>} for \process{}es.

Finally, the \lhe{} struct is a wrapping type for both \event{}s and \process{}es, enabling transposition between the two formats. When only one of the two formats is loaded, transposition can be achieved through the member functions \codett{transpose()}, and when both exist one can be uses as basis to override the other with the overloaded member function \codett{transpose(std::string source)} --- or, to explicitly decide, one can call the member functions \codett{events\_to\_processes()} or \codett{processes\_to\_events()}. When transposing from \codett{event}s to \codett{process}es, the boolean member \codett{filter\_processes} determines whether to transpose directly from the \codett{event} data or from the reordered and possibly filtered \codett{event\_view}.

By default, the \codett{lhe::transpose()} function will sort events into individual subprocesses by their external partons, and will filter the data to that relevant to those external partons. The former can be changed by defining an \codett{event\_hash\_fn} either using the \codett{eventSorter} type detailed in \cref{sec:rex_wrappers} or with a custom hash as discussed in \cref{sec:rex_fundamentals}, while the latter can be changed directly by changing the boolean \codett{filter\_processes} member.

Additionally, the \lhe{} type has a \codett{header} member stored as \codett{std::any} since the header itself is not defined in the LHE format --- aside from the fact that reweighting information is stored in an \codett{<initrwgt>} child node. Using the default reader, the \codett{header} is stored as a shared pointer to a \rex{} format \codett{xmlNode}, i.e. identically to the node in the read LHE file, allowing access and modification directly in the XML format. The \codett{<initrwgt>} is also modified online when appending new weights to the \lhe{} type, assuming that the stored \codett{header} is of type \codett{std::shared\_ptr<xmlNode>}, meaning a rewritten reweighted LHE file will automatically account for new weights not just in the individual \event{}s but also in the \codett{header}. Additionally, just like \event{}s and \process{}es, \lhe{} can store arbitrary information in the \codett{std::unordered\_map<std::string,std::any>} \codett{extra}.

Besides event-level data stored in OO and SoA formats, the \lhe{} types also stores process characterisation data according to the LHE standard. Like event characterisation data, this can be accessed by reference using several different access functions, as detailed in \cref{tab:lhe_data}.

\begin{table}[t]
    \centering
    \begin{tabular}{c|c|c}
        \textbf{Data} &  \textbf{Type} & \textbf{Access functions} \\ \hline
        Beam IDs & \codett{long int[2]} & \codett{idBmUP()}, \codett{idBm()}\\
        Beam energies & \codett{double[2]} & \codett{eBmUP()}, \codett{eBm()} \\
        pdf group IDs & \codett{short int[2]} & \codett{pdfGUP()}, \codett{pdfG()}\\
        pdf set IDs & \codett{long int[2]} & \codett{pdfSUP()}, \codett{pdfS()}\\
        Weight scheme & \codett{short int} & \codett{idWgtUP()}, \codett{idWgt()} \\
        No. processes & \codett{unsigned short} & \codett{nProcUP()}, \codett{nProc()} \\
        Cross section(s) & \double{} & \codett{xSecUP()}, \codett{xSec()}\\
        Error(s) & \double{} & \begin{tabular}{c}
              \codett{xSecErrUP()},\\ \codett{xSecErr()}
        \end{tabular} \\
        Max weight(s) & \double{} & \codett{xMaxUP()}, \codett{xMax()}\\
        Process ID(s) & \codett{long int} & \codett{lProcUP()}, \codett{lProc()}
    \end{tabular}
    \caption{Access functions for \codett{<init>} node data in the LHE standard. Note that the process-specific information (\codett{xSec}, \codett{xSecErr}, \codett{xMax}, and \codett{lProc}) are properties of individual processes, an arbitrary amount of which can be stored in a single LHE file, and as such these variables are actually stored as \codett{std::vector}s of the corresponding type, indexed according to the order they show up in the LHE file.}
    \label{tab:lhe_data}
\end{table}

Reading LHE files can be done through the free function \codett{load\_lhef()}, which is overloaded to take either \codett{std::istream} objects or file paths given as \codett{std::string}s as arguments, returning the loaded \lhe{} object. Similarly, \lhe{} objects can be written to disk using the free function \codett{write\_lhef()} which is similarly overloaded to write either to a \codett{std::ostream} or to a file path given by an argument \codett{std::string}, although for \codett{write\_lhef()} the first argument must be the loaded \lhe{} object.

Besides LHE files, has simple support for the SLHA format for model parameters \cite{Boos:2001cv}. The \codett{slha} class provides a simple dictionary-like storage container for named blocks of parameter types, with each parameter in each block defined by an \codett{int} ID and a \double{} value. These parameters can be accessed and modified through the functions:
\begin{lstlisting}[language=rex]
double get(const std::string &block,
   int index, double fallback = 0.0);
void set(const std::string &block,
   int index, double value);
\end{lstlisting}
with \codett{fallback} the value returned if the given parameter cannot be found. \codett{slha} objects can be constructed from \codett{std::istream}s or \codett{std::string}s using the free functions \codett{to\_slha(...)}, or loaded from disk with the free function \codett{load\_slha(const std::string \&filename)}. Note that \rex{} lacks support for named parameters, meaning all parameters must be defined by both a block name and an ID.

As a sidenote, as mentioned above \rex{} comes shipped with \codett{xmlDoc} and \codett{xmlNode} types, internally used for parsing the XML-based LHE format. The \rex{} XML parser was developed for two reasons:
\begin{enumerate}
    \item Avoiding external dependencies; \rex{} and \tearex{} are intended to be entirely self-contained in order to ensure minimal issues when including them in other software, as well as avoiding long-term stability issues regarding different versions of other packages.
    \item Optimised usage; generic XML parsers have very different design goals than what is needed for \rex{}, and consequently very different optimisation targets; the LHE format has well-defined conventions and minimal hierarchical structure, making generic XML parsing excessive and generally bloated\footnote{Before the first internal XML parser was developed, several external XML parsers were used as placeholders. All tested ones either had issues regarding memory consumption or load speed, which the \rex{} parser overcomes due to assumptions about the LHE format.}. 
\end{enumerate}
Consequently, \rex{} comes shipped with a small, simple XML parser which may not adhere entirely to the full XML standard, but should a user want to interact with LHE files in this format (or XML files in general) it is possible. Unlike the \lhe{} type, \rex{} does not support direct file loading into the \codett{xmlNode} format --- instead, an XML file needs to be loaded into a \codett{std::string}, after which it can be loaded into \codett{std::shared\_ptr<xmlNode>} using
\begin{lstlisting}[language=rex]
std::shared_ptr<REX::xmlNode> node 
   = REX::xmlNode::parse(raw);
\end{lstlisting}
at which point the XML file (or node) is accessible.

More technical details on the \codett{xmlNode} type are provided in \cref{sec:rex_fundamentals}; for the remainder of this section, we will limit ourselves to listing some of the relevant functionality of the class. First, XML node content can be read using the following member functions:
\begin{itemize}
    \item \codett{std::string\_view name()}: Returns a view of exclusively the node name, excluding any attributes stored in the start tag.
    \item \codett{std::string\_view content()}: Returns a view of the full node (including children) \textit{excluding} the start and end tags.
    \item \codett{std::string\_view full()}: Returns a view of the full node (including children) \textit{including} the start and end tags.
\end{itemize}

XML attributes are stored as a minimal struct \codett{Attr} with members \codett{Attr::name\_view} and \codett{Attr::value\_view}, both stored as string views and accessible through the member functions \codett{name()} and \codett{value()}, respectively. The attributes of an \codett{xmlNode} can be accessed through the member function
\begin{lstlisting}[language=rex]
const std::vector<Attr> &attrs();
\end{lstlisting}
which provides const read-only access. Attributes can be added and modified using the \codett{xmlNode} member functions
\begin{lstlisting}[language=rex]
void add_attr(std::string name_, 
   std::string value_);
bool set_attr(std::string_view name_,
   std::string value_);
\end{lstlisting}
where the former adds a new attribute with the given name and value, while the latter sets an existing attribute to the new value. The returned bool from \codett{set\_attr} is \codett{true} if successful and \codett{false} if no attribute with the given name is found.

Children of an \codett{xmlNode} are stored as a vector of (shared pointers to) \codett{xmlNode}s. This vector can be accessed through the \codett{children()} member function, although more extensive child treatment is possible using the following functions:
\begin{itemize}
    \item \codett{void add_child(std::shared_ptr<xmlNode>} \codett{child)}: Appends the given child to the end of the parent node.
    \item \codett{bool has_child(std::string_view name_)}: Checks if any children have a given name.
    \item \codett{std::shared\_ptr<xmlNode> get\_child} \codett{(std::string\_view name_)}: Returns the first child with the given name, returning a \codett{nullptr} on failure to find any.
    \item \codett{std::vector<std::shared\_ptr<xmlNode>>} \codett{get\_children(std::string\_view name_)}: Returns a vector of all children with name \codett{name_}.
    \item \codett{bool remove\_child(...)}: Overloaded function which suppresses a given child from being written, returning \codett{true} on success and \codett{false} if the child could not be found. Argument can be either \codett{size\_t} giving the position in the vector of children, \pointer{xmlNode} giving the address of the child, or \codett{std::string\_view} giving the name of the child. The final overload will only suppress the first child with the corresponding name, and thus has limited use for files with many identically named nodes.
    \item \codett{bool replace\_child(size\_t anchor,} \codett{std::shared\_ptr<xmlNode> child)}: Suppresses the child at the given position in the vector of children and replaces it for writing. Can also be called with \codett{std::string\_view name} instead of \codett{size\_t}, replacing the first child with the given name. Returns \codett{true} on success and \codett{false} on failure to find the given child.
\end{itemize}
This list is not comprehensive, but should provide enough functionality for typical use cases. The full public functionality can be read from the header \codett{Rex.h}.

\subsubsection{Functionality wrappers}
\label{sec:rex_wrappers}

In order to support generic functionality without forcing users to create all beyond-default functionality from scratch, \rex{} comes with a plethora of methods for constructing custom versions of most functionality, such as comparators and sorters for \event{}s, generic extra information for \event{} and \lhe{} objects, and generic writers and readers from and to the \lhe{} data format. 

Starting with the generic \codett{extra} information stored in \event{}s and \process{}es, it is a member variable of type \codett{std::unordered\_map} which maps a name given as \codett{std::string} to a value of type \codett{std::any} (for \event{}) or \codett{std::vector<std::any>} (for \process{}). For \process{} objects this map needs to be accessed directly as a member variable \codett{process.extra}, but the \event{} type has templated access operators
\begin{lstlisting}[language=rex]
template <typename T>
void set(const std::string&, T)
{...}
T &get(const std::string&)
{...}
const T &get(const std::string&) const
{...}
\end{lstlisting}
with the \codett{get} functions internally handling the \codett{std::any\_cast<T\&>} before returning the given object, as well as throwing a bad-any-cast error if called with the wrong type \codett{T} or an out-of-range error if no element with the given name is found. To safely check whether \codett{extra} has a given entry, the boolean member function \codett{bool has(const std::string\&)} will test existence without trying to access the entry. The \codett{initNode} type, from which the \lhe{} type inherits, has identical functionality for more generic non-event level information.

Sorting \event{}s into \process{}es in \rex{} is rather simple, although there are several types necessary to build up the sorting infrastructure. Although completely custom operators can be provided (as detailed in \cref{sec:rex_fundamentals}), custom sorting schemes can also be created with built-in \rex{} routines by creating event comparators using the \codett{eventComparatorConfig} type, which when combined with a set of events can create a boolean pass/fail filter for an input event using the \evbelongs{} type; by combining several \evbelongs{} objects a custom sorting/hash function can be created with the \codett{eventSorter} type.

The \codett{eventComparatorConfig} struct is made up of boolean \codett{compare\_...} members defining whether a particular trait of the LHE standard should be compared to determine whether two \event{}s are ``equal'' or not -- each of which comes with self-returning setters for all the access names provided for \event{}s in \cref{tab:rex_event_data,tab:rex_parton_data} --- as well as tolerances for how much all data of type \double{} may differ (relatively) to determine equality. Additionally, a \codett{std::set<int>} member \codett{status\_filter} allows defining the relevant values of \codett{iStUP} for which partons to compare such that only partons whose status is included in \codett{status\_filter} are included for \event{} comparisons and any whose status is omitted are ignored (unless it is empty, in which case no filtering is applied). Although tolerances for each \double{} value are stored separately, no setters are defined for these other than the generic self-returning \codett{set\_tolerance(double)}, which sets all tolerances to the given value; should varied tolerances be required, consult the header \codett{Rex.h} to see what these variables are named. The self-returning \codett{set\_status\_filter}, however, is overloaded to support calls with \codett{std::vector<int>}, \codett{std::set<int>}, or any generic \codett{Args...}; in the last case, it is assumed the arguments can trivially be converted into elements of \codett{std::set<int>}.

Once an \codett{eventComparatorConfig} has been customised, the resulting comparator can be accessed using the \codett{make\_comparator()} (or \codett{make\_const\_comparator()} member functions. This returns an \codett{event\_equal\_fn}, a boolean function type which takes two \event{}s as input and returns whether the \event{}s are equivalent under the corresponding comparison, i.e. essentially a custom \codett{operator==} (see \cref{sec:rex_fundamentals} for more details on local and global equivalence comparisons). The default \codett{eventComparatorConfig} setup will compare all parton statuses, PDG codes, and masses, although the default \event{} sorter only compares the PDG codes of external legs, creating \evbelongs{} objects for each set of external legs.

The \evbelongs{} type is simple: It is equipped with a vector of \event{}s and an \codett{event\_equal\_fn} function pointer (which can be set by users) and can test whether an input \event{} matches any of its \event{}s with respect to the comparator through the \codett{belongs(event\&)} member function, which can also be called through \codett{operator()} (i.e. for an \evbelongs{} \codett{eb} and an \event{} \codett{e}, \codett{bool eb(e)}). This allows for relatively free-form pass/fail tests on \event{}s, and provides the basis for the default hashing structure used to sort \event{}s in the \lhe{} struct.

A function of an \event{} returning a bool is of the type \codett{event\_bool\_fn} per \rex{} standards, and \evbelongs{} objects can emit their resulting \codett{event\_bool\_fn} through the \codett{get\_event\_bool()} member function. Sorting functions for \event{}s are defined as
\begin{lstlisting}[language=rex]
using event_hash_fn = 
   std::function<size_t(event&)>
\end{lstlisting}
which essentially is any function mapping an \event{} to \codett{size\_t}. Such hash functions can be created from \evbelongs{} objects with the \codett{eventSorter} type, which is even simpler than the \evbelongs{} type: It consists of a vector of \evbelongs{} objects, and when its member function \codett{position(event\&)} is called it returns the first index to which the \event{} the call to \codett{eventBelongs::belongs} returns \codett{true}. On a failure to map the \event{} to any index, it returns \codett{npos}. Similarly to how \evbelongs{} can provide functions for corresponding \codett{event\_bool\_fn}s, the \codett{eventSorter} struct has the member function \codett{get\_hash()} which returns a function pointer to its hashing function. The \lhe{} type has a member variable \codett{eventSorter sorter} which can be set using the self-returning setter \codett{set\_sorter} which is used at runtime to determine the splitting of owned \event{}s when transposing to the \process{} format. \codett{lhe::set\_sorter} can also be called with a generic \codett{event\_equal\_fn} to automatically generate an \codett{eventSorter} based on currently owned events.

With the internal treatment of LHE data described, I/O routines remain to be detailed. \rex{} comes equipped with two generic templated types \codett{lheReader} and \codett{lheWriter} which, as the names suggest, can be used to convert the \lhe{} type to and from generic data formats. Although internally these types have some significant differences, for an end-user the experience is largely equivalent: The \codett{lheReader} needs to be supplied with conversion functions from the template types \codett{EventRaw} and \codett{InitRaw} to the corresponding \rex{} types \event{} and \codett{initNode} --- as well as optionally a function mapping a template \codett{HeaderRaw} to \codett{std::any} for the generic \codett{lhe.header} object --- and vice versa for \codett{lheWriter}. For both classes, these translators are defined as
\begin{lstlisting}[language=rex]
using InitTx = std::function
   <initNode(const InitRaw&>;
using EventTx = std::function
   <event(const EventRaw&)>;
using HeaderTx= std::function
   <std::any(const HeaderRaw&)>;
\end{lstlisting}
and the other way around for \codett{lheWriter}. Additionally, \codett{EventTx} has surrounding helpers to support \codett{std::function}s returning not only an \event{} object, but also \codett{std::shared\_ptr<event>} or \codett{std::unique\_ptr<event>}.

These types can be constructed using the constructors
\begin{lstlisting}[language=rex]
lheReader(initTx in_tx, EventTx ev_tx);
lheWriter(initTx in_tx, EventTx ev_tx);
\end{lstlisting}
with \codett{HeaderTx} as an optional additional argument which will be ignored unless set, and we reiterate that the translator functions have opposite directionality for the two types. Alternatively, both types have self-returning setters \codett{set\_init\_translator}, \codett{set\_event\_translator}, and \codett{set\_header\_translator}.

\codett{lheReader} has the member function \codett{read}, which takes as input an \codett{InitRaw} and an \codett{EventRange}, as well as an optional \codett{std::optional<HeaderRaw>}. \codett{EventRange} must be an iterable object such that the following loop is well-defined:
\begin{lstlisting}[language=rex]
for (const auto &er : events_raw)
   {
      evts.emplace_back(event_tx_(er));
      ...
   }
\end{lstlisting}
e.g. \codett{std::vector<EventRaw>} or a similar type. Alternatively, the free function
\begin{lstlisting}[language=rex]
lhe read_lhe(const InitRaw &in_raw,
   const EventRange &ev_raw,
   InitTx in_tx,
   EventTx ev_tx,
   std::optional<HeaderRaw>
      header_raw = std::nullopt,
   HeaderTx head_tx = nullopt)
\end{lstlisting}
can be called to handle all the details automatically and just return the resulting \lhe{} object. Similarly, \codett{lheWriter} has the \codett{to\_raw(const lhe \&doc)} member function which returns an \codett{lheRaw} object storing the resulting information (detailed below); alternatively, the free function
\begin{lstlisting}[language=rex]
lheRaw write_lhe(const lhe &doc,
   InitTx in_tx,
   EventTx ev_tx,
   HeaderTx head_tx = nullopt)
\end{lstlisting}
can be called to handle the intricacies.

The \codett{lheRaw} struct is a minimal storage container for raw LHE information, having only three members: \codett{InitRaw init}, \codett{std::vector<EventRaw> event}, and the optional \codett{HeaderRawOpt header} (which must be of type \codett{std::optional<...>}. As of \rex{} version 1.0.0 \codett{lheReader} does not support reading using the templated \codett{lheRaw} type, nor does it or \codett{lheWriter} support the automatic I/O of generic types using function pointers for \codett{std::istream} and \codett{std::ostream}, but should there be interest in such user-end simplifications, they could be implemented for future versions.

\subsubsection{Fundamental types}
\label{sec:rex_fundamentals}

Although the C++ \codett{std::array} and \codett{std::vector} types provide contiguous storage for fixed size (\codett{array}) and dynamic size (\codett{vector}) sequential type containers, as of the C++17 standard there is no container for dynamically sized containers of fixed size containers\footnote{While e.g. a vector of arrays necessitates the arrays to be contiguous and the elements of the arrays to be contiguous, the elements of sequential arrays are not necessarily adjacent in memory as the arrays may have start and end padding. While the \codett{std::mdspan} introduced in C++23 does not have this restriction, it is not yet supported by most compilers.}. A container that supports multidimensional indexing while ensuring memory contiguity is especially important when trying to optimise HEP code, as many (and typically the most important) quantities are defined in terms of four-vectors and two-/four-spinors. This becomes especially important when optimising code for hardware acceleration using SIMD instructions or SIMT machines.

To treat this deficiency, three templated fundamental types are defined in \rex{}: The fixed size array \codett{arrN}, reference-like fixed size access objects \codett{arrNRef}, and the dynamically sized vector of arrays \codett{vecArrN}. These are templated with respect to the underlying object type \codett{typename T} and the array dimensionality \codett{size\_t N} with explicit library-side instantiations for \codett{T}$\,\in\{$\codett{short, long, int, float, double}$\}$ and \codett{N}$\,\in\{2,3,4\}$. Additionally, aliases for \codett{N}$\,\in\{2,3,4\}$ are provided: 
\begin{lstlisting}[language=rex]
using arr2<T> = arrN<T,2>;
using arr3<T> = arrN<T,3>;
using arr4<T> = arrN<T,4>;
\end{lstlisting}
and similarly for \codett{arrNRef} and \codett{vecArrN}.

\codett{arrN<T,N>} is a wrapper for a C-style array with type \codett{T} and size \codett{N}, with some additional functionality to circumvent the common pitfalls of using \codett{T[N]}. However, \codett{arrNRef<T,N>} is just a non-owning proxy which can be accesses in the same manner as \codett{arrN} --- at its core, it is simply a pointer \codett{T} \pointer{q} with reference access to the \codett{N} memory locations starting at the location \pointer{q} --- allowing for \codett{arrN}-like access to data stored in a separate container. Finally, \codett{vecArrN<T,N>} is just \codett{std::vector<T>} equipped with a custom iterator \codett{nStrideIter<T,N>} such that e.g. \codett{vecArrN<double,4>[0,1,2,...,M]} returns an \codett{arrNRef<double,4>} object with \pointer{q} pointing to the underlying element at  \codett{std::vector<double>[0,4,8,...,4}$\times$\codett{M]} etc. Similarly, operators relating to the size of a \codett{vecArrN} object are multiplied and divided by \codett{N} for access, reservation, sizes, and so on. However, the underlying vector can also be accessed by reference with the method \codett{vecArrN::flat\_vector()}, which is e.g. how \process{} momenta are passed to the scattering amplitude routines in \madtrex{} (detailed further in \cref{sec:madtrex}).

Note that \codett{arrN} and its derived types are designed for trivial SIMD alignment specifically for real numbers -- while one can construct an \codett{arrN<std::complex>} trivially, the interweaved real and imaginary parts make it unsuited for SIMD operations. However, it would be relatively simple to write a complex-like class \codett{cArrN<T,N>} consisting of two separate \codett{arrN<T,N>} objects corresponding to the real and imaginary parts of the array with explicitly defined elementary operations \codett{+,-,*,/}. While such a class is not provided with \rex{} version 1.0.0, it could be implemented in a future version alongside derived \codett{cArrNRef<T,N>} \codett{vecCArrN<T,N>} types should there be interest for it.

Furthermore, \codett{arrN} and \codett{vecArrN} only provide 1- and 2-dimensional storage and access, making matrix and tensor multiplication non-trivial. We do note that that \codett{arrN} has a method \codett{arrN::dot(const arrN\& other)} allowing for generic Euclidean dot products between \codett{arrN} objects. This can be used to implement matrix multiplication between \codett{arrN} and \codett{vecArrN} objects, but again, such developments are left for future work.

Although we leave the description of the \event{} and \process{} types for later, we note here that \codett{arrN} and derived types are used to store not just momenta (\codett{arr4<double>}, but also particle mothers (\codett{arr2<short int>}) and colour flow (\codett{arr2<short int>})\footnote{\codett{arrN} is also used to store beam IDs, beam energies, pdf groups, and pdf sets, but here there is no particular reason to prefer \codett{arrN} over \codett{std::array}.}.

Aside from the storage types, there are some function type aliases that are relevant for in-depth \rex{} usage. The first are event comparator types:
\begin{lstlisting}[language=rex]
using event_equal_fn = 
   std::function<bool(event &, event &)>;
using cevent_equal_fn = 
  std::function<bool
  (const event &, const event &)>;
\end{lstlisting}
which can be used as generic equality comparators for event objects and serve as the foundation of the boolean \evbelongs{} event testers and hashing \codett{eventSorter} event sorters. The reason for both mutable and const versions of this type is simple: Mutable comparators can sort event partons online while doing the comparison. While this feature is not used inside \rex{}, it could be used to optimise the process of sorting \event{}s into individual \process{}es. The default \codett{operator==} for \event{} objects exclusively compares the (unordered) PDG codes of the external legs, but this can be changed globally using the function
\begin{lstlisting}[language=rex]
void set_event_comparator
   (cevent_equal_fn fn);
\end{lstlisting} 
Internally, the \codett{operator==} function has access to a \codett{std::shared\_ptr<cevent\_equal\_fn>} which shares scope with a \codett{std::mutex} that is only accessible from \codett{operator==} and \codett{set\_event\_comparator}. Note that the global comparator must be of the const type.

The other two function types relevant for low-level configuration are
\begin{lstlisting}[language=rex]
using event_bool_fn = 
   std::function<bool(event &)>;
using event_hash_fn = 
   std::function<size_t(event &)>;
\end{lstlisting}
both of which also have corresponding const \codett{cevent\_...} types. The former is a generic pass/fail type for \event{} objects, although internally in \rex{} it serves no purpose other than acting as an intermediate type between the comparators above and the hash type just shown. The \codett{event\_hash\_fn} type, however, is a fundamental part of \rex{}, as it is used when transposing \lhe{} objects between the \event{} and \process{} formats --- the \lhe{} struct has an \codett{event\_hash\_fn} member, and before creating the \process{} objects all events are sorted based on the indexing provided by this hash function. Specifically, each unique hash value will be mapped to a unique process ordered as per the order the events appear in the \lhe{} data\footnote{I.e. if we have unique hashes 0 and 1, the \process{} objects may be ordered with events corresponding to hash 1 first if the first \event{} object has hash 1. This avoids segmentation faults when sorting processes, but means custom hash functions may not end up corresponding to the ordering of the resulting processes.}, with the caveat that events with hash \codett{REX::npos = (size\_t)-1} will be mapped to the very last \process{} corresponding to ``unsorted'' events. Custom hashes can be provided to the \lhe{} object using the self-returning member function \codett{set\_hash(event\_hash\_fn hash)}, where we note that \codett{cevent\_hash\_fn} can be automatically converted to \codett{event\_hash\_fn} (although the opposite conversion is impossible).

One final intricacy beyond the scope of typical usage is the XML parser supplied with \rex{}, although the specifics here are unlikely to be interesting for power users. To give a brief description, the \codett{xmlNode} class is a tree-like structure of pointers between mother nodes and children with a shared loaded data storage in \codett{std::string} format, to which all the individual nodes only have access through \codett{std::string\_view}s of their respective data. Children can be added, removed, and replaced, with new nodes (or node attributes) owned by the child in question, while any data left unmodified is kept as \codett{std::string\_view}s of the original \codett{std::string}.

We omit more extensive details on XML treatment as \rex{} is not primarily intended to for XML utility, although we reiterate that \rex{} XML handling is designed with respect to primarily reading and secondarily writing LHE format files and that it may not perfectly fulfil the XML standard when handling generic XML files nor be particularly efficient in handling more complex node hierarchies than the almost linear LHE format.

\subsection{Use case illustrations}
\label{sec:rex_usecase}

\begin{algorithm*}
\begin{lstlisting}[language=rex]
REX::eventComparatorConfig comp1,comp2;
// Comparators for all external legs and only final-state particles
comp1.set_status_filter({-1,1}).set_pdg(true).set_mass(true);
comp2.set_status_filter({1}).set_pdg(true).set_mass(true);
auto exeternalComp = comp1.make_const_comparator();
auto finalComp = comp2.make_const_comparator();

REX::event ev1, ev2;
ev1.set_n(4).set_pdg({21,21,6,-6}).set_status({-1,-1,1,1}).set_mass({0,0,173,173});
ev2.set_n(4).set_pdg({2,-2,6,-6}).set_status({-1,-1,1,1}).set_mass({0,0,173,173});
assert( !externalComp(ev1,ev2) );
assert( finalComp(ev1,ev2) );

// Changing global comparators
REX::set_event_comparator( externalComp );
assert( !(ev1 == ev2) );
REX::set_event_comparator( finalComp );
assert( ev1 == ev2 );
\end{lstlisting}
\caption{Illustration of event comparisons using the \codett{eventComparatorConfig} type to easily create generic comparison operators which can either be called locally or set for global comparisons.}\label{code:rex_ev_comp}
\end{algorithm*}

In this section, we intend to illustrate some uses for the functionality mentioned above to show how simple \rex{} is to use for writing new software or for unifying a format for existing software. Starting with some elementary uses, LHE files can be read, sorted and transposed, and written by
\begin{lstlisting}[language=rex]
REX::lhe file = REX::load_lhef(inpath);
file.transpose();
std::ofstream out(outpath);
file.print(out);
\end{lstlisting}
where the default event sorting algorithm was used, comparing the PDG codes of external partons. Alternatively, an illustration of the more generic comparison capabilities is shown in \cref{code:rex_ev_comp}. \rex{} internally always uses explicitly defined comparison operators, leaving users free to define and utilise them however they want within their codebase --- of course, noting that other types need their local \codett{event\_bool\_fn}s defined as well. Note that the function \codett{externalComp} shown in \cref{code:rex_ev_comp} is equivalent to the default event comparison operators used throughout \rex{}.

With the simplicity of constructing event comparators illustrated, it follows that it is just as simple to create the boolean pass/fail tests provided by the \codett{REX::eventBelongs} type:
\begin{lstlisting}[language=rex]
std::vector<REX::event> es = ...
event_equal_fn cmp = ...
auto belong = REX::eventBelongs(es,cmp);
\end{lstlisting}
and with that we can test whether a given \event{} fits our particular conditions formulated in terms of other \event{}s and specific fields of comparison with these. This immediately extends to the creation of \codett{event\_hash\_fn}s using the \codett{REX::eventSorter} type as
\begin{lstlisting}[language=rex]
std::vector<REX::eventBelongs> belong =
   ...
REX::eventSorter hash(belong);
\end{lstlisting}
and an \event{} can now be hashed through the member function \codett{REX::eventSorter::position}, and vectors of \event{}s can be hashed at once using the member function \codett{REX::eventSorter::sort}. Note that a failed hash will always return \codett{REX::npos}.

Of course, \codett{eventSorter}s can also be constructed from generic \codett{(c)event\_bool\_fn}s, and similarly, the \codett{event\_hash\_fn} used when transposing \event{}s in a \codett{REX::lhe} object to the \codett{REX::process} type can be set explicitly using the \codett{lhe::set\_hash} member function for more generic uses, but the simpler application of \codett{eventBelongs} objects should suffice for most use cases.

One additional note for event sorting is the ordering of \codett{particle}s in an \event{}. As mentioned above, individual \codett{particle}s are given as views of the corresponding data row stored in an \event{}, and these can either be accessed directly from the \event{} --- in which case the \codett{particle}s are ordered according to the underlying data storage --- or through a \codett{REX::event\_view}, which just masks the \codett{REX::event::operator[]} through the member vector \codett{REX::event::indices}, i.e. (somewhat simplified)
\begin{lstlisting}[language=rex]
particle event_view::operator[](size_t i)
{ return this->evt[indices[i]]; }
\end{lstlisting}
where \codett{indices} will default to \codett{\{0,1,...\}} unless explicitly set beforehand. Setting these indices is done through the \event{} member function \codett{set\_indices}, which is overloaded to take as input either a vector of \codett{size\_t}s, \textit{or} another event to be indexed with respect to\footnote{By indexing an \event{} (\codett{orig}) with respect to another (\codett{oth}), we mean setting the indices such that when \codett{orig} is accessed through the \codett{event\_view}, \codett{particle}s will have the same order as they would in the data structure of \codett{oth}, the event we indexed with respect to. This indexing only considers the PDG codes and statuses of partons, and will ignore partons that do not appear in \codett{oth}, i.e. their position will not be included in the vector of indices.}. While for most usecases \codett{event::indices} need to be set explicitly, \rex{} has one access point where they are set automatically: When calling \codett{eventBelongs::belongs} with a non-constant \event{}, if the \event{} succeeds its \codett{set\_indices} member function will be called with the \event{} it was compared to:
\begin{lstlisting}[language=rex]
bool eventBelongs::belongs(event &e){
   ...
   for(auto ev : this->events){
      if(this->comparator(*ev,e)){
         e.set_indices(*ev);
         return true;
      }
   }
   return false;
}
\end{lstlisting}
which is how \lhe{} objects are sorted by default, which, when combined with the default member setting \codett{filter\_processes = true} means transposed \process{} objects unless otherwise specified will be filtered to the particles used for event comparison and each \event{}'s data ordered with respect to the events used in sorting. This particular feature is important for consideration when using the \process{} type as input to (data-parallel) functions where it is assumed that each contiguous data set is ordered identically, which is assumed and used in \madtrex{} (elaborated on in \cref{sec:madtrex}).

\begin{algorithm*}
\begin{multicols}{2}
\begin{lstlisting}[language=rex]
using xmlReader = lheReader<
  std::shared_ptr<xmlNode>, 
  std::shared_ptr<xmlNode>,
  std::shared_ptr<xmlNode>
>;

using xmlWriter = lheWriter<
  std::shared_ptr<xmlNode>,
  std::shared_ptr<xmlNode>,
  std::optional<std::shared_ptr<xmlNode>>
>;

using xmlRaw = lheRaw<
  std::shared_ptr<xmlNode>,
  std::shared_ptr<xmlNode>,
  std::optional<std::shared_ptr<xmlNode>>
>;

const xmlReader &xml_reader(){
   static const xmlReader b{&xml_to_init, 
      &xml_to_event, &xml_to_any};
   return b;
}

const xmlWriter &xml_writer(){
   static const xmlWriter t{&init_to_xml,
      &event_to_xml, &header_to_xml};
return t;
}

xmlRaw to_xml_raw(const lhe &doc){
   return xml_writer().to_raw(doc);
}

\end{lstlisting}
\end{multicols}
\caption{Usage illustration for the \codett{REX::lheReader} and \codett{REX::lheWriter} types, using function pointers to the corresponding translators \codett{xml\_to\_TYPE} for types \codett{REX::initNode}, \codett{REX::event}, and \codett{std::any} (with \codett{std::any} used as a generic container for the \codett{<header>} node in the LHE format), and vice versa for translators \codett{TYPE\_to\_xml}.}\label{code:rex_xmlparser}
\end{algorithm*}

The final functionality important to consider is support for generic I/O, using wrappers for ``translator functions'' between the \event{} and \codett{initNode} formats and some arbitrary alternate types \codett{InitRaw} and \codett{EventRaw}. Of course, it is entirely possible to create a generic \codett{REX::lhe} constructor from an arbitrary data format, but our intention here is to provide a minimal constructor for any arbitrary data format.

The templated \codett{REX::lheReader} and \codett{REX::lheWriter} classes are from a user-side perspective incredibly similar --- they both use two to three ``translator'' function pointers in order to construct a translator to-or-from the \codett{REX::lhe} type and a generic input/output format.

First, we consider \codett{lheReader}. It has three template arguments,
\begin{lstlisting}[language=rex]
template <class InitRaw, class EventRaw,
   class HeaderRaw = std::monostate>
   class lheReader{
   public:
   using InitTx = std::function<initNode
      (const InitRaw &)>;
   using EventTx = std::function<
      std::shared_ptr<event>(
         const EventRaw &)>;
   using HeaderTx = std::function<
   std::any(const HeaderRaw &)>;
...}
\end{lstlisting}
where the \codett{HeaderRaw} type is not necessary but allows for handling of generic data containers for LHE headers. The \codett{lheReader} type has two explicit constructors:
\begin{lstlisting}[language=rex]
explicit lheReader(InitTx init_tx,
   EventTx event_tx)
{ set_init_translator(
     std::move(init_tx));
  set_event_translator(
     std::move(event_tx));
}
\end{lstlisting}
which only treats the mandatory translators for the \event{} and \codett{initNode} types. The second constructor additionally handles an optional translator from a generic \codett{HeaderRaw} type to the \codett{std::any} type used to store the \texttt{<header>} LHE node:
\begin{lstlisting}[language=rex]
template <class InitTx, 
   class EventTx, class HeaderTx>
lheReader(InitTx init_tx, 
   EventTx event_tx, HeaderTx header_tx)
{  set_init_translator(
       std::move(init_tx));
   set_event_translator(
       std::move(event_tx));
   set_header_translator(
      std::move(header_tx));
}
\end{lstlisting}
i.e. \codett{lheReader}s need translators for \event{}s and \codett{initNode}s, and optionally also \codett{header}s, where the resulting type for \codett{header} is \codett{std::any}. Note that while \codett{EventTx} is defined in terms of the type \codett{std::shared\_ptr<REX::event>} there are surrounding helpers converting raw \event{}s or \codett{std::unique\_ptr}s to \event{}s to the \codett{std::shared\_ptr} format used in \lhe{}, meaning the user-provided \codett{EventTx} does not need to provide a shared pointer directly (although we of course suggest using shared pointers where applicable to ensure object continuity). Alternatively, \codett{lheReader} like most \rex{} types has self-returning setters \codett{set_x_translator},  for \texttt{x} $\in$ \{\texttt{init,event,header}\}. Or, to minimise type interfacing, the templated free function \codett{read\_lhe} can be used:
\begin{lstlisting}[language=rex]
template <class InitRaw, class EventRange,
  class HeaderRaw = std::monostate,
  class InitTx, class EventTx,
  class HeaderTx = std::nullptr_t>
lhe read_lhe(const InitRaw &init_raw,
  const EventRange &events_raw,
  InitTx init_tx, EventTx event_tx,
  std::optional<HeaderRaw> header_raw = 
     std::nullopt,
  HeaderTx header_tx = nullptr,
  bool filter_processes = false)
{using EventRawT = typename std::decay<
   decltype(*std::begin(
   events_raw))>::type;
 lheReader<InitRaw, EventRawT, HeaderRaw>
   b;
 ...
 return b.read(
   init_raw, events_raw, header_raw);
}
\end{lstlisting}
which handles all the intricacies of \lhe{} construction, only necessitating the relevant translators and object sets to construct \lhe{} objects. The \codett{lheWriter} type is similar, with the caveat that it has an intermediate return type \codett{lheRaw} to store \codett{initNode} data, \event{} data, and optionally \codett{header} data,
\begin{lstlisting}[language=rex]
template <class InitRaw, class EventRaw, 
   class HeaderRawOpt>
struct lheRaw
{
   InitRaw init;
   std::vector<EventRaw> events;
   HeaderRawOpt header;
};
\end{lstlisting}
and is just a minimal storage container for the raw data (assuming that events are stored in an object-oriented format).

With this in mind, \codett{lheWriter} is defined almost identically to \codett{lheReader},
\begin{lstlisting}[language=rex]
template <
   class InitRaw, class EventRaw,
   class HeaderRaw = std::monostate>
class lheWriter
{  public:
   using InitTx = std::function<InitRaw(
     const initNode &)>;
   using EventTx = std::function<
     EventRaw(event &)>;
   using HeaderTx = std::function<
     HeaderRaw(const std::any &)>;
   using result_t = lheRaw<
     InitRaw, EventRaw, HeaderRaw>;
  lheWriter(InitTx init_tx,
     EventTx event_tx, 
     HeaderTx header_tx = HeaderTx{})
   : init_fn_(std::move(init_tx)), 
     event_fn_(std::move(event_tx)), 
     header_fn_(std::move(header_tx)) {}
...}
\end{lstlisting}
and identically to \codett{lheWriter} it has self-returning setters \codett{set\_init\_translator}, \codett{set\_event\_translator}, as well as an optional \codett{set\_header\_translator}. And, again, there is the free function \codett{write\_lhe}, although this  one has the form
\begin{lstlisting}[language=rex]
template <class InitRaw,
  class EventRange,
  class EventRawT = typename
  std::decay<decltype(*std::begin(
    std::declval<EventRange>()))>::type,
  class HeaderRaw = std::monostate,
  class InitTx, class EventTx,
  class HeaderTx = std::nullptr_t>
lheRaw<InitRaw, EventRawT, HeaderRaw> 
  write_lhe(lhe &doc,
    InitTx init_tx,
    EventTx event_tx,
    HeaderTx header_tx = nullptr)
\end{lstlisting}
where the only significant change to \codett{read\_lhe} is that the leading argument is now a singular \codett{REX::lhe} object rather than all the raw objects. Although \rex{} internally uses direct conversions from singular XML nodes to \rex{} types and writes directly to \codett{std::ostream}, it comes shipped with implementations for the \codett{REX::xmlNode} format to provide illustrations. For reference, these are shown in \cref{code:rex_xmlparser}, where the exact usage of these types are provided for the \codett{xmlNode} types without details regarding the internal structure of the \codett{xmlNode} itself. Essentially, given a function that turns e.g. a generic \codett{EventRaw} into a \codett{REX::event}, a translator can trivially be constructed and called using this function and the source objects.

\subsection{Benchmarks}
\label{sec:rex_benchmarks}

While it is difficult to profile the full extent of a wide-ranging library such as \rex{}, we can still benchmark some of its standard functionality. For the purpose of providing a reasonable showcase of what we expect to be typical usecases for \rex{}, we here present the event throughput of some standard functionality for a generic workflow: throughputs (in terms of events per second) for reading and sorting LHE files. We will analyse this using the standard XML-based LHE file format --- which \rex{} comes shipped with parsers for --- to provide a benchmark for \rex{}' efficiency.

\begin{figure}
    \centering
    \includegraphics[width=0.95\linewidth]{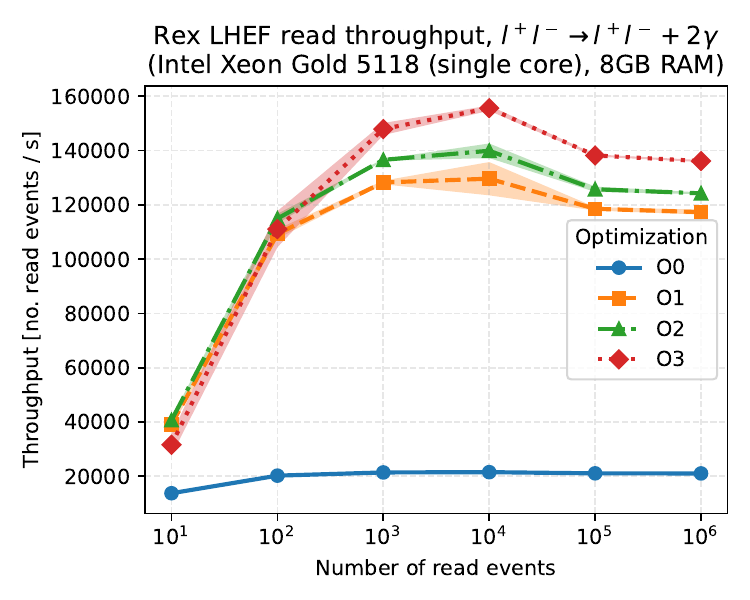}
    \caption{Read throughput for an electroweak LHE sample using \rex{} as a function of file size in terms of number of events for optimisation levels \oz{} through \oth{} using \codett{g++} version 13.2.0. Each point gives an average throughput from 100 measurements, with standard deviations highlighted. To ensure only the library functionality itself was measured, the benchmark executable was compiled with no optimisations. It is clear that the most significant speed-up comes from \oo{} optimisation, although \otw{} and particularly \oth{} do provide additional load speed-up (compare with \cref{fig:rex_readtp_pp}, where \oth{} has no significant speed-up compared to \otw{}.) The dip at 100 000 events coincides with the corresponding LHE file exceeding the 16.5 MB L3 cache of the Intel Xeon Gold 5118 \cite{Intel5118Specs}.}
    \label{fig:rex_readtp_ll}
\end{figure}

\begin{figure}
    \centering
    \includegraphics[width=0.95\linewidth]{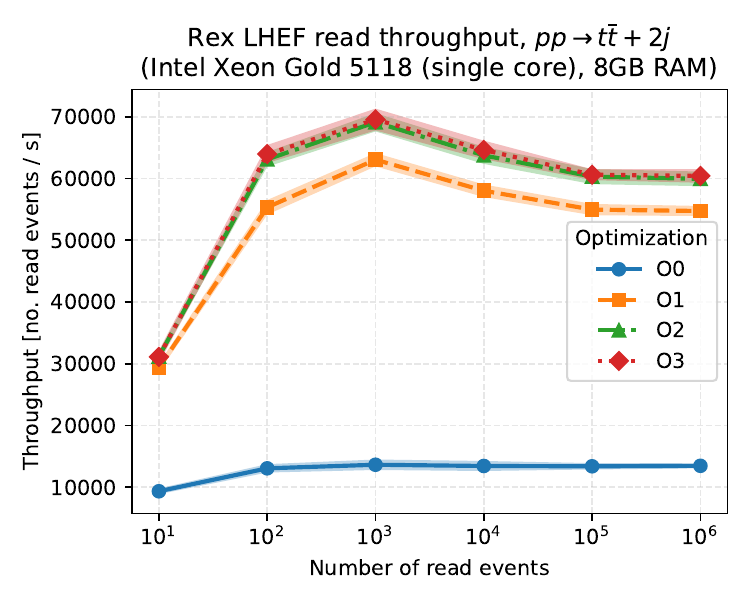}
    \caption{Read throughput for a QCD LHE sample using \rex{} as a function of file size in terms of number of events for optimisation levels \oz{} through \oth{} using \codett{g++} version 13.2.0. Each point gives an average throughput from 100 measurements, with standard deviations highlighted. To ensure only the library functionality itself was measured, the benchmark executable was compiled with no optimisations. It is clear that the most significant speed-up comes from \oo{} optimisation, although \otw{} does provide additional load speed-up while \oth{} does not appear to have any significant impact (compare with \cref{fig:rex_readtp_ll}, where \oth{} has significant speed-up compared to \otw{}.) Although the throughput dip is more gradual than in \cref{fig:rex_readtp_ll}, the plateau is once again reached at 100 000 events, which also for these samples is when the LHE file size exceeds the 16.5 MB L3 cache of the Intel Xeon Gold 5118 \cite{Intel5118Specs}}
    \label{fig:rex_readtp_pp}
\end{figure}

Starting with read throughput: since the XML-based default format in \rex{} needs to parse any additional node data, we will test both electroweak samples (whose events contain only information belonging to the LHE standard) and QCD samples (whose events may contain additional information regarding e.g., the event-specific pdf scale) generated with \mg{}. These measurements are provided in \cref{fig:rex_readtp_ll,fig:rex_readtp_pp}, where the function \codett{REX::load\_lhef} was timed in total 100 times for each file for each tested level of compiler optimisation (from none with \oz{} to maximal with \oth{}) applied to the \rex{} library compilation. For all tests, the benchmark executable was compiled without optimisation to ensure times were representative of only \rex{} functionality.

\Cref{fig:rex_readtp_ll,fig:rex_readtp_pp} provide several interesting insights, especially when compared. First and foremost, at least \oo{} optimisation is necessary for \rex{} to reach its potential --- for both sample sets, \oo{} provides a roughly factor 6 speed-up compared to no compiler optimisation, with \oth{} only being marginally faster than \oo{}. Additionally, for both samples a throughput plateau is reached at the samples with $10^5$ events, which for both sets coincides with the size of the loaded LHE file exceeding the 16.5 MB L3 cache of the Intel Xeon Gold 5118 tests were run on at 86.5 MB and 110.0 MB, respectively\footnote{\codett{REX::load\_lhef} streams the loaded file rather than reading it directly into memory, but native data types are only marginally smaller than the plaintext LHE format. A back-of-the-envelope calculation suggests \event{}s with only data given by the LHE standard should at most be $\sim50\%$ smaller than the corresponding plaintext, excluding any padding or other surrounding infrastructure. Some preliminary tests on $\Delta$RSS suggest that a \codett{REX::lhe} object is similarly sized to the corresponding LHE file, reinforcing this conclusion.}. More notable are the significantly different throughputs measured in \cref{fig:rex_readtp_ll,fig:rex_readtp_pp}: already at 100 events, the EW sample is read twice as fast as the QCD sample. As far as \rex{} is concerned, the only significant difference between these files is the additional \codett{<mgrwt>} child node each event has, storing renormalisation and pdf information. The likeliest source of slowdown are the string operations in appending these children to the member \codett{REX::event::extra} (detailed in \cref{sec:rex_wrappers}), although more extensive tests are left for future development.

\begin{figure}
    \centering
    \includegraphics[width=0.95\linewidth]{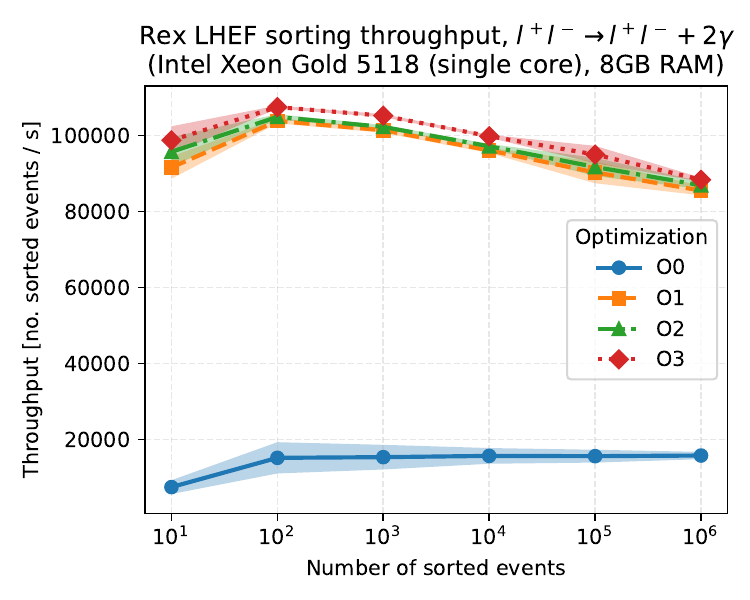}
    \caption{Read throughput for an electroweak LHE sample using \rex{} as a function of file size in terms of number of events for optimisation levels \oz{} through \oth{} using \codett{g++} version 13.2.0. Each point gives an average throughput from 100 measurements, with standard deviations highlighted. To ensure only the library functionality itself was measured, the benchmark executable was compiled with no optimisations. It is clear that the most significant speed-up comes from \oo{} optimisation.}
    \label{fig:rex_sorttp_ll}
\end{figure}

\begin{figure}
    \centering
    \includegraphics[width=0.95\linewidth]{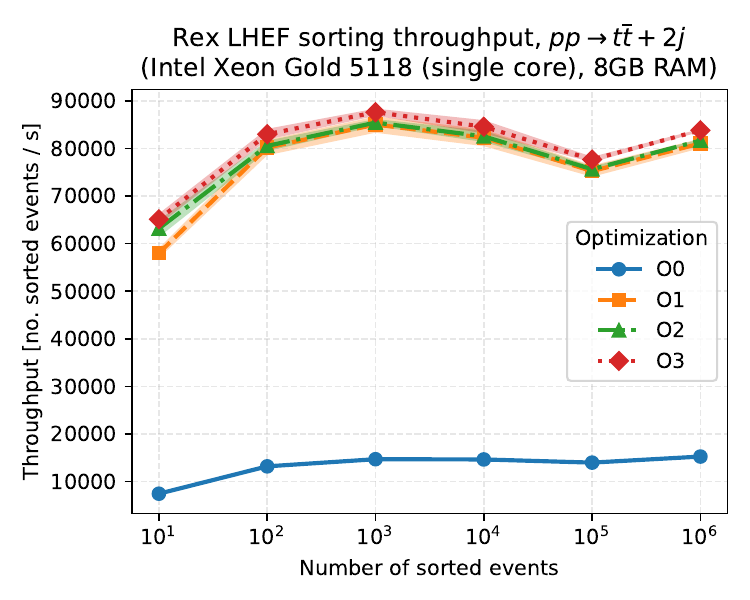}
    \caption{Read throughput for a QCD LHE sample using \rex{} as a function of file size in terms of number of events for optimisation levels \oz{} through \oth{} using \codett{g++} version 13.2.0. Each point gives an average throughput from 100 measurements, with standard deviations highlighted. To ensure only the library functionality itself was measured, the benchmark executable was compiled with no optimisations. It is clear that the most significant speed-up comes from \oo{} optimisation.}
    \label{fig:rex_sorttp_pp}
\end{figure}

Next, we turn to event sorting. Specifically, for an already loaded \codett{REX::lhe} we measure the runtime of the function call
\begin{lstlisting}[language=rex]
lhefile.transpose();
\end{lstlisting}
i.e. the time taken to both sort the \codett{REX::event} members and then extract their information into \codett{REX::process} objects. Furthermore, we use the default sorting method where the \codett{REX::lhe} object creates a \codett{REX::eventSorter} online, comparing external partons of events and appending the current external legs to the sorter if the sorter does not recognise this particular configuration. The measured throughputs, again taken as the mean of 100 measurements with standard deviations highlighted, are shown in \cref{fig:rex_sorttp_ll,fig:rex_sorttp_pp}.

Once any initial overhead is overcome, the expected complexity scaling of sorting LHE files (just as for reading) is $\mathcal{O}(\#\text{events})$ which should result in a roughly constant event throughput. This aligns well with \cref{fig:rex_sorttp_ll,fig:rex_sorttp_pp}. However, there are some interesting points of consideration: first, in comparison to \cref{fig:rex_readtp_ll,fig:rex_readtp_pp} almost all compiler optimisation here comes from \codett{-O1} optimisation, with no notable difference between \codett{-O1} and \codett{-O2} past the minimal samples with 10 events; additionally, the differences in throughput between the sample sets are small. One final observation is the different behaviour between the sample sets as functions of the number of events: the EW sample has its peak throughput early, before decreasing slightly with increasing sample sizes; on the other hand, the QCD sample has a throughput increase between the samples with $10^5$ and $10^6$ events.

The differences between \cref{fig:rex_sorttp_ll,fig:rex_sorttp_pp}, and \cref{fig:rex_readtp_ll,fig:rex_readtp_pp} are unsurprising: unlike read, which involves file streaming, string manipulation, type conversions etc., \codett{REX::eventBelongs}-based sorting only involves object comparisons and copies, leaving little to optimise beyond memory management and function call ordering. Furthermore, the read overhead seen for QCD samples in \cref{fig:rex_readtp_pp} is driven by the additional string manipulation when loading events' XML child nodes --- by the time sorting occurs, these are already stored as \codett{std::shared\_ptr<REX::xmlNode>}s owned by their events and when transposing to the \codett{REX::process} format only a pointer is copied rather than the full string content, making the overhead between EW and QCD samples far less significant in sorting than for reading.

One final point of interest is the differing behaviour between \cref{fig:rex_sorttp_ll,fig:rex_sorttp_pp}. Although we do not know the origin of the increased throughput going to the million events sample in \cref{fig:rex_sorttp_pp}, it is not surprising that the two figures behave slightly differently. To see this, let us consider the complexity scaling of event sorting more in-depth: we have noted that we should see complexity scaling as $\sim\mathcal{O}(\#\text{events})$ (i.e. throughput should change minimally as the number of events per sample increases), but the leading constant comes from several origins. In particular, two:
\begin{align}
    t_{\text{total}} \simeq t_{\text{sort}} + t_{\text{transpose}},
\end{align}
i.e. the two runtime costs are the event hashing and the transposition of event data to \codett{REX::process} objects, both of which clearly should be linear in time. However, recall the default hashing mechanism for \codett{REX::lhe} objects: events are sorted into groups based on their unordered external legs, with initial- and final-state particles separated. Particularly, this means that the time taken when sorting a sample depends on the number of distinct parton configurations within that sample, even if the following transposition will differ minimally in runtime. For reference, the samples used for \cref{fig:rex_sorttp_ll} have 6 distinct parton configurations, all of which are sampled already for the 100-event sample. On the other hand, the samples in \cref{fig:rex_sorttp_pp} have 65 distinct parton configurations which in the unweighted samples tested are sampled at very different rates --- in fact, only the $10^6$-event sample actually contains all 65 configurations. With this in mind, differing throughputs in \cref{fig:rex_sorttp_pp} may just be luck with respect to the ordering of events in the LHE file; if the file happens to start with less common configurations, the dominant ones will have to go through more comparisons before having their hash determined.

\section{Tensorial event adaption}
\label{sec:tearex}

The \texttt{tensorial event adaption with Rex} library --- \tearex{} --- is an extension to \rex{} adding support for completely generic parton-level event reweighting using the SIMD- and SIMT-friendly \process{} data format for sets of events. Simply, \tearex{} adds a \rwgtr{} type inheriting from the \codett{REX::lhe} type which in addition to \lhe{} members also owns a vector of \procrwgt{}s; a type defined to perform event-level reweighting for singular \process{} objects and automatically appending them to the \event{}s owned by the relevant \rwgtr{} type object. Additionally, \tearex{} comes shipped with support for so-called ``matrix element reweighting'' where a given event is reweighted to different values of model parameters assuming model parameters are stored as SLHA cards on disk and scattering amplitudes can be evaluated from \codett{REX::process} objects. Further details are provided in \cref{sec:madtrex}, which primarily uses this implementation for its SIMD- and SIMT-enabled leading order reweighting.

For reference, parton-level event weights in HEP are defined by \cite{Mattelaer:2016gcx}
\begin{align}
    w = f_1(x_1,\mu_F) f_2(x_2,mu_F) \; \msq{} \; \Omega_{PS},
\end{align}
with $f_i$ the pdf evaluated at Bjorken fraction $x_i$ and factorisation scale $\mu_F$, $\msq{}$ the absolute scattering amplitude squared of the event at the given phase space point, and $\Omega_{PS}$ the phase space measure of the corresponding event. Notably, these contributions all factorise, meaning that evaluating the resulting event weights from changing one of them can be calculated without having to re-evaluate the others. As an example, consider parameter reweighting where some physics model parameter entering the scattering amplitude $\m{}$ is changed, yielding
\begin{align}
    w' &=  f_1(x_1,\mu_F) f_2(x_2,mu_F) \; \mprsq{} \; \Omega_{PS}\\
       &= \left( \prod_{i=1,2} f_i(x_i,\mu_F) \right) \; \msq{} \frac{\mprsq{}}{\mprsq{}} \; \Omega_{PS}\\
       &= \frac{\mprsq{}}{\msq{}} w,
\end{align}
i.e. weights in a new model can be defined in terms of weights in the original one as long as the new model weights are non-zero only in a subspace of the phase space of the original model. Since different stages of HEP event simulation generally factorise (e.g. hard scattering events are disparate from hadronisation are disparate from detector simulation) this implies event samples simulated in one model can be repurposed for a different one under certain mathematical restrictions by only re-evaluating the scattering amplitude of the underlying hard scattering process.

A user manual for \tearex{} is provided in \cref{sec:tearex_manual} --- which gives extensive details for how to use \tearex{} for parton-level event reweighting --- and some implementation use case codes are shown in \cref{sec:tearex_usecase}.

\subsection{Manual}
\label{sec:tearex_manual}

\tearex{} is a small extension to \rex{} which adds types for appending new event weights to existing \lhe{} objects using completely generic ``reweighting functions'' (\weightor{}s), by which we mean any function acting on a \process{} object and returning a (shared pointer) to a vector of \double{}s;
\begin{lstlisting}[language=rex]
using weightor = std::function<
   std::shared_ptr<std::vector<double>>
   (process &)>;
\end{lstlisting}
which are intended to evaluate (arbitrary) resulting weights for all events stored in a \process{} object. If reweighting parameters are global, \tearex{} also supports ``reweighting iterators'' run between calls to \weightor{}s and can thus safely modify global data without impeding the \weightor{} calls,
\begin{lstlisting}[language=rex]
using iterator = std::function<bool()>;
\end{lstlisting}
which can e.g. be used to reweight an entire event set for differing model parameters as is done in \madtrex{} (cf. \cref{sec:madtrex}).

This manual will be split into three sections: \cref{sec:tearex_default,sec:tearex_generic} illustrate how to use \tearex{} functionality to implement event reweighting using the user-friendly helper functions provided by \rex{} as well as completely generally, respectively; then \cref{sec:tearex_slha} showcases the specific SLHA parameter reweighting implementation shipped with \tearex{} for use in \madtrex{}.

\subsubsection{Default usage}
\label{sec:tearex_default}

The primary functionality of \tearex{} is provided by the \procrwgt{} and \rwgtr{} types --- plus the helper \codett{threadPool} type managing multithreading between individual \process{}es --- which handle the LHE-level handling of weight information treatment when reweighting events. The key assumptions made for \tearex{} are:
\begin{itemize}
    \item Each \process{} object is exclusively well-defined, i.e., no event belongs to multiple subprocesses.
    \item Individual \process{}es are independent algorithmically and can be reweighted simultaneously.
    \item Reweighting routines are symmetric with respect to particle ordering\footnote{This can be overcome with more specific parton sorting in individual events, using details from \cref{sec:rex_fundamentals,sec:tearex_generic}. However, the default sorting routines provided by \rex{} only consider particle status (whether it is incoming, outgoing, or internal) and thus cannot treat anything more complex than differentiating between initial- and final-state particles.}.
\end{itemize}

Starting with the \procrwgt{} struct, it is an independent type not inheriting from any other types, primarily to avoid double-counting information (as the \rwgtr{} type inherits from the \codett{REX::lhe} type). Simply put, it has a boolean pass/fail filter to determine whether \event{}s belong to its corresponding \process{} (given by a \codett{REX::event\_bool\_fn} or \codett{REX::eventBelongs} object), a \weightor{} \codett{normaliser} defining the original weights of its events, a vector of \weightor{}s to perform the relevant reweighting, and a shared pointer to its corresponding \process{} which is intended to be simultaneously owned by the \procrwgt{} and the surrounding \rwgtr{}. The \procrwgt{} type has constructors
\begin{lstlisting}[language=rex]
procReweightor(weightor rwgt_fn);
procReweightor(weightor rwgt_fn,
   REX::eventBelongs selector);
procReweightor(
   std::vector<weightor> rwgts);
procReweightor(std::vector<weightor> rws,
   REX::eventBelongs selector);
procReweightor(std::vector<weightor> rws,
   REX::eventBelongs selector,
   weightor normaliser);
\end{lstlisting}
where each constructor including \evbelongs{} has an overloaded corresponding constructor using \codett{std::shared\_ptr<REX::eventBelongs>} which we advise users to prefer when plausible\footnote{Due to the intricate ownership tree for most type provided by \rex{}, it is plausible for objects to unintentionally fall out of scope --- this is why shared pointers are so extensively used in the suite, aside from the contexts where it is important for objects to have shared ownership (such as \process{}es and \lhe{}).}. Note that the order of parton-level variables by default are sorted according to the order of said particles in the \event{}s making up the \evbelongs{} object when transposing from the \event{} to the \process{} format, as shown in \cref{sec:rex_usecase}. Alternatively, instead of an \evbelongs{} object an arbitrary pass/fail filter can be set using the \codett{event\_bool\_fn} type, which is elaborated on in \cref{sec:tearex_generic}.

Each \procrwgt{} has a shared pointer to the \process{} object it is meant to reweight (\codett{std::shared\_ptr<REX::process> proc}) and a vector of corresponding \weightor{}s (\codett{std::vector<weightor> reweight\_functions}) --- function pointers to each corresponding reweighting function. These contain the central functionality for reweighting an individual \process{}, performed by the member function \codett{evaluate(size\_t amp)}:
\begin{lstlisting}[language=rex]
void procReweightor::evaluate(size_t amp)
{...
auto wgts = this->reweight_functions
   [amp](*this->process)
...}
\end{lstlisting}
with some additional surrounding checks irrelevant for this description. Additionally, unless explicitly defined at the level of \procrwgt{}, which \process{} belongs to which \procrwgt{} object is defined by the running \rwgtr{} and will be ignored here; for now, we assume there is a clear one-to-one relationship between \process{}es and \procrwgt{}s.

At launch, the member function \codett{void initialise()} should be called. After checking that the \procrwgt{} has a \process{}, \codett{initialise()} checks whether the \procrwgt{} has a member function pointer \codett{normaliser(process\&)}; if it does not, it will set its \codett{normaliser} to be the first available \weightor{}, and if no \weightor{}s are available it will only return zero-valued weights. This is particularly important for hash misses, as elaborated on below, but for now it should be seen as a safety fallback if a \procrwgt{} is initialised improperly.

Once \codett{initialise()} has ensured that the \procrwgt{} has a \codett{normaliser} and a \process{}, it will run
\begin{lstlisting}[language=rex]
auto norm = this->normaliser(this->proc);
std::transform(norm->begin(),
   norm->end(), norm->begin(),
   [](double val){
   return (val == 0.0) ?
   0.0 : 1.0 / val; });
this->normalisation = 
   *(REX::vec_elem_multi<double>
   (*norm, this->proc->weight());
\end{lstlisting}
i.e. for e.g. parameter reweighting it will set its normalisation factor to be $w/\msq{}$, leaving all new weights to be normalised by a multiplication with the corresponding normalisation factor. Then, with this in mind, once the \procrwgt{} object gets the go-ahead to store a given reweighting iteration it will run
\begin{lstlisting}[language=rex]
for(auto &wgts : this->backlog{
    this->proc->append_wgts(
    *(REX::vec_elem_multi<double>
    wgts, this->normalisation));}
\end{lstlisting}
where \codett{backlog} is a member vector storing new weights between the call to \codett{evaluate()} and the end of the current reweighting iteration.

\procrwgt{} members can be set using the self-returning setters
\begin{lstlisting}[language=rex]
procReweightor &set_event_checker
   (REX::eventBelongs checker);
procReweightor &set_normaliser
   (weightor normaliser);
procReweightor &set_reweight_functions
   (weightor rwgt);
procReweightor &set_reweight_functions
   (std::vector<weightor> rwgts);
procReweightor &add_reweight_function
   (weightor rwgt);
procReweightor &set_process
   (std::shared_ptr<REX::process> pr);
\end{lstlisting}
although we reiterate that the \process{} is generally intended to be set at the level of the \rwgtr{} type rather than the \procrwgt{}.

As mentioned above, the \rwgtr{} type inherits from the \codett{REX::lhe} type and adds additional functionality to sort owned \event{}s using owned \procrwgt{}s; run the process-specific reweighting functions; iterate over global states; appending resulting weights to the corresponding \event{}s; and calculating the resulting reweighted cross section $\sigma'$ and corresponding cross section error $\Delta\sigma'$. Aside from inherited constructors and direct constructors from the \lhe{} type, \rwgtr{} has the additional constructors
\begin{lstlisting}[language=rex]
reweightor(lhe &&mother,
   std::vector<procReweightor> rws);
reweightor(lhe &&mother,
   std::vector<procReweightor> rws,
   std::vector<iterator> iters);
\end{lstlisting}
and corresponding constructors with \codett{std::shared\_ptr<procReweightor>} replacing \procrwgt{} (which we reiterate are to be preferred), as well as corresponding copy constructors for the \lhe{} object. In addition to the reweighting \iterator{}s shown in the constructors above, the \rwgtr{} type has two additional free \iterator{}s \texttt{initialise} and \codett{finalise} intended to set and reset the global state to what it should be for and after the full reweighting procedure, respectively. As for most other \rex{} and \tearex{} types, \rwgtr{} has self-returning setters for most of its members:
\begin{lstlisting}[language=rex]
reweightor &set_reweightors(
   std::vector<procReweightor> rws);
reweightor &add_reweightor(
   procReweightor &rw);
reweightor &set_initialise(
   iterator init);
reweightor &set_finalise(
   iterator fin);
reweightor &set_iterators(const
   std::vector<iterator> &iters);
reweightor &add_iterator(
   const iterator &iter);
\end{lstlisting}
and corresponding setters using \codett{std::shared\_ptr<procReweightor>}. One additional member that can be set as above are tags for individual reweighting runs, stored in the \codett{std::vector<std::string> launch\_names} which also has a setter and an element adder as above.

Internal details on the \rwgtr{} types are given in \cref{sec:tearex_generic}, but the general algorithm which is run by calling the member function \codett{void run()} is:
\begin{enumerate}
    \item Call \texttt{initialise()}
    \item Construct an \codett{event\_hash\_fn} from \procrwgt{}s' owned \evbelongs{}
    \item Sort \event{}s with the constructed hash
    \begin{itemize}
        \item If there are unsorted events, add a \procrwgt{} which returns weights zero
    \end{itemize}
    \item Return the resulting \process{}es to the corresponding \procrwgt{}s which run their member \codett{normaliser} on their \process{}
    \item Set up the \codett{threadPool} of workers
    \item For each owned \iterator{}:
    \begin{enumerate}
        \item Run the \iterator{}
        \item For each \procrwgt{}, submit a job to the \codett{threadPool}
        \item Once a job is launched; run each reweighting \weightor{} owned by that \procrwgt{}
        \item Wait for all jobs to finish
        \item Append returned weights to their \process{}es
    \end{enumerate}
    \item Transpose weights from \process{}es to \event{}s
    \item Call \codett{finalise()}
    \item If there are \codett{launch\_names}, add them to the common list of weight tags
    \item Calculate reweighted cross sections and run error propagation for them
\end{enumerate}
While this list is long, it is simple; in particular, most of the details are regarding the setup and wind down, while the individual reweighting iterations are simple. Note that reweighted cross sections and cross section errors are calculated despite not being stored as part of the LHE standard --- using the default \rex{} writer they will not be written to disk, but they can be accessed in-software through the corresponding \rwgtr{} members
\begin{lstlisting}[language=rex]
std::vector<double> rwgt_xSec;
std::vector<double> rwft_xErr;
\end{lstlisting}
which are given more detail in \cref{sec:tearex_generic}. Again, as a reminder, once a \rwgtr{} object has been set up using the constructors or setters above, the full reweighting procedure with all steps described above are performed by just calling the member function \codett{run()}. The reweighted LHE file can then be written to disk using the default LHE writer or a custom writer as shown in \cref{sec:rex_manual}.

\subsubsection{Generic reweighting}
\label{sec:tearex_generic}

While \cref{sec:tearex_default} describes most of the functionality and practical details of \tearex{}, there are some more involved possible uses, as well as some internal details, that may be of interest for more complicated implementations. These concern generic event hashing, details on the multithreading helper \codett{threadPool}, the mathematical details of reweighted cross sections, and the possible implementation of single-event reweighting using the object-oriented \event{} type rather than the SoA \process{} type.

First, the \procrwgt{} type has overloaded constructors and setters for the \codett{event\_bool\_fn} type rather than the \evbelongs{} type. These allow for a generic way of defining which \event{}s belong to which \rwgtr{} in a less restricted format than that provided by the \evbelongs{} type (although obviously requiring more end-user programming), i.e.
\begin{lstlisting}[language=rex]
procReweightor(weightor rwgt_fn,
   REX::event_bool_fn selector);
procReweightor(std::vector<weightor> rws,
   REX::event_bool_fn selector);
procReweightor(std::vector<weightor> rws,
   REX::event_bool_fn selector,
   weightor normaliser);
procReweightor &set_event_checker
   (REX::event_bool_fn selector);
\end{lstlisting}
However, this excludes the automatic parton indexing applied when using the \evbelongs{} type and consequently means the parton ordering in the transposed \process{}es will be identical to that stored in the original \event{} unless new indices are explicitly set in the \codett{selector}.

Let us now turn to the \codett{threadPool} type, used in \tearex{} to schedule and launch individual \procrwgt{}s across separate CPU threads. The type itself is a minimal wrapper for the \codett{std::vector<std::thread>} member \codett{workers\_}, with size defined at construction,
\begin{lstlisting}[language=rex]
threadPool::threadPool(unsigned int t){
   workers_.reserve(t);
   for (unsigned i = 0; i < t; ++i){
     workers_.emplace_back(
     ...
     );}
}
\end{lstlisting}
where the omitted section is just a lambda function for grabbing tasks from the member \codett{std::queue<std::function<void()>> q\_} and error handling. To set up jobs, assuming the given tasks are stored in a vector of \codett{std::function<void()> jobs}, is as simple as
\begin{lstlisting}[language=rex]
std::vector<std::function<void()> 
   jobs = {...};
threadPool pool(t);
pool.begin_batch();
for(auto job : jobs){
   pool.enqueue(job);
}
pool.wait_batch();
\end{lstlisting}
and the program will then wait until the batch is finished before continuing. By default, the \codett{pool} used by the \rwgtr{} type is constructed with the number of threads available in the current context as provided by \codett{std::thread::hardware\_concurrency()}, but can be set explicitly using the \rwgtr{} member \codett{unsigned int pool\_threads}.

Next, we turn towards the mathematical details of reweighted cross sections and error propagation. As mentioned, these are not part of the LHE standard itself (as they can be calculated from the cross section and the individual event weights), but \tearex{} nevertheless provides the functionality to evaluate them directly through the reweighting interface. Reweighted cross sections are given as
\begin{align}
    \sigma' = C\sum_i w'_i,
\end{align}
with $C$ the same normalisation as the original event sample. This trivially extends to arbitrary observables as long as said observables are independent of the form of the hard scattering process, although \tearex{} currently only treats cross sections. Error propagation is performed assuming Gaussian behaviour as \cite{Mattelaer:2016gcx}
\begin{align}
    \Delta \,\sigma' = \Delta \, \sigma\; \cdot \left( \frac{1}{N} \sum_{i=1}^N \frac{w'_i}{w_i}  \right) + \sigma \; \cdot \; \text{std}(w'),
\end{align}
where by std we refer to the standard deviation of a variable. If error propagation for any reweighted cross section fails --- by e.g. returning infinity, NaN, or non-positive --- a warning is raised without throwing an error and the error is estimated as the original error $\Delta \sigma$ multiplied by the ratio of the reweighted and original cross sections (conservatively always taking whichever ratio is greater than one).

One final point of consideration is the possible implementation of event-by-event reweighting using the OO \event{} data format. While \tearex{} does not natively support this directly, to motivate the usage of the SoA \process{} format for HEP software, it is possible to implement it indirectly by noting that the definition of the \weightor{} type,
\begin{lstlisting}[language=rex]
using weightor = std::function<
   std::shared_ptr<std::vector<double>>
   (process&)>;
\end{lstlisting}
does not enforce the usage of any specific members of the \process{} type, only the usage of the \process{} type itself. As mentioned in \cref{sec:rex}, the \process{} objects owned by \lhe{} objects have shared access to their corresponding \event{}s through vectors of shared pointers to \event{}s, i.e.
\begin{lstlisting}[language=rex]
std::vector<std::shared_ptr<event>> 
   lhe.processes[i]->events = 
   lhe.sorted_events[i];
\end{lstlisting}
from which it immediately follows that event-by-event reweighting can be performed in \tearex{} by wrapping a loop over individual \event{} reweighting calls in a \codett{weightor}, e.g.
\begin{lstlisting}[language=rex]
double foo(const REX::event &ev){...}
REX::tea::weightor fooWrap = [](
  std::shared_ptr<std::vector<double>> pr)
  {auto wgts = std::make_shared
      <std::vector<double>>({});
   for(auto e : pr.events){
      wgts->push_back(foo(*e));}
   return wgts;
  };
\end{lstlisting}
and while not directly supported nor recommended, this is a possible use of \tearex{}.

\subsubsection{SLHA parameter reweighting}
\label{sec:tearex_slha}

As part of \tearex{} an explicit reweighting implementation is provided, used as the basis for SLHA parameter reweighting in \madtrex{}. A detailed description of it is provided below to illustrate practically how to implement a reweighting application using the \tearex{} library.

The SLHA parameter reweighting is defined using just two new types besides those already implemented in \rex{} and \tearex{}: \rwgtslha{}, which generates SLHA parameter cards for reweighted parameters and writes them to disk; and \paramrwgt{}, a small \rwgtr{} child type with ownership of a \rwgtslha{} object to generate its \iterator{}s. How parameter reweighting is implemented in \tearex{} works is algorithmically simple: There exists some externally provided scattering amplitude functions in the \weightor{} format, and these \weightor{}s read model parameters from a parameter card on disk. The \rwgtslha{} object is provided with a reweighting card with the format
\begin{lstlisting}
launch rwgt_name=run1
set BLOCK_NAME PARAM_ID VALUE
set BLOCK_NAME PARAM_ID VALUE
launch rwgt_name=run2
set BLOCK_NAME PARAM_ID VALUE...
\end{lstlisting}
as well as an original SLHA parameter card to modify the parameter values in. Inheriting from the \codett{REX::slha} type, these commands are easily translated to \iterator{}s which overwrite the parameter card with one containing the new parameters, as well as an \texttt{initialise} function copying the original card to a safe location and a \codett{finalise} function moving the original card back into position. Aside from surrounding safety and sanity checks, this is implemented as
\begin{lstlisting}[language=rex]
bool write_rwgt_card(size_t idx){
for(auto [key,val] : cards[idx].blocks){
   for(auto [p_id, p_val] : val.params){
      original.add_param(key, p_id, 
         this->get(key, p_id));
      this->set(key, p_id, p_val);}
   }
this->write(std::ofstream(card_path));
}
\end{lstlisting}
with some simplifications made for easier reading, noting that the member \codett{original} is an owned \codett{REX::slha} object just storing the original values of modified parameters to ensure the original value of reweighted parameters are reset before reweighting new ones. In short, \codett{write\_rwgt\_card} writes SLHA parameter cards identical to the original card save for the parameters modified for that particular reweighting iteration (as given by the launch commands mentioned above). From this, the reweighting \iterator{}s are provided by a simple function call
\begin{lstlisting}[language=rex]
std::vector<iterator> card_writers(){
std::vector<iterator> writers;
for(size_t i=0; i < cards.size(); ++i){
   writers.push_back([this,i]){
      return write_rwgt_card(i);
   };}
return writers;
}
\end{lstlisting}
and from this the entire foundation for a \rwgtr{} is presented; \tearex{} provides \process{}-specific reweightors using \weightor{}s, but in this case a global variable (i.e. model parameters) is modified using \iterator{}s before running the same \weightor{} routines which are dependent on the global state set by the \iterator{}s. In fact, the few additional members of the \paramrwgt{} type besides those inherited from \rwgtr{} are only there to interface directly with the \rwgtslha{} type to minimise necessary interfacing in implementations: \paramrwgt{} has a single unique member function, which passes a reweighting card in the format above to its \rwgtslha{} member and then initiates its own \texttt{initialise}, \codett{finalise}, and \iterator{}s to those provided by the \rwgtslha{}:
\begin{lstlisting}[language=rexstar]
void read_slha_rwgt(std::istream &slha,
   std::istream &rwgt){
   card_iter = rwgt_slha::create
      (slha, rwgt);
   initialise = [&](){
   return card_iter.move_param_card();};
   finalise = [&](){
   return card_iter.remove_param_card();
   };
   iterators = card_iter.
      get_card_writers();
   ...
}
\end{lstlisting}
with some additional lines afterwards passing information about the reweighting onto the \lhe{} context for writing. Fundamentally, though, \paramrwgt{} is just a \rwgtr{} with \iterator{}s provided by the commands given in a reweighting card in the SLHA format; besides that, all it needs are \weightor{}s alongside corresponding \event{} sorters.

\subsection{Use case illustrations}
\label{sec:tearex_usecase}

\begin{algorithm*}
\begin{multicols}{2}
\begin{lstlisting}[language=rex]
using namespace REX;
std::vector<event> gg, gq, qq;
std::vector<int> qs = 
  {1,-1,2,-2,3,-3,4,-4};
  
for(size_t q1 = 0; q1 < qs.size(); ++q1){
 for(size_t q2 = q1; q2 < qs.size(); ++q2)
 {event ev_qq(2).set_status({-1,-1})
    .set_pdg({qs[q1],qs[q2]});
  qq.push_back(ev); }
 event ev_gq(2).set_status({-1,-1})
   .set_pdg({21,qs[q1]});
 gq.push_back(ev_gq);
}

gg.push_back(event(2).set_status({-1,-1})
  .set_pdg({21,21}));

auto comp = eventComparatorConfig()
  .set_pdg(true).set_mass(true)
  .set_status_filter({-1})
  .make_comparator();

eventBelongs quark(qq, comp);
eventBelongs mixed(gq, comp);
eventBelongs gluon(gg, comp);

std::vector<tea::procReweightor> rwgtrs;

tea::weightor quark_pdf_fn = ...
tea::weightor mixed_pdf_fn = ...
tea::weightor gluon_pdf_fn = ...

rwgtrs.push_back(tea::procReweightor(
   quark_pdf_fn, quark);
rwgtrs.push_back(tea::procReweightor(
   mixed_pdf_fn, mixed);
rwgtrs.push_back(tea::procReweightor(
   gluon_pdf_fn, gluon);

lhe file_to_reweight = ...

tea:reweightor rwgt_runner(
   file_to_reweight, rwgtrs);

rwgt_runner.run();
\end{lstlisting}
\end{multicols}
\caption{General outline for a program to run pdf reweighting using \rex{}, including explicit definitions of sorting operators to split the sample into events with two initial-state quarks, two initial-state gluons, or one of each. The actual pdf sets have been omitted, but would be provided through the \codett{tea::weightor} objects listed in the code, and could either be implemented as global functions with \codett{tea::iterator}s changing them globally between iterations or instead as vectors of \codett{tea::weightor}s with one element per pdf set. The only assumption here is that the order of the initial-state particles is unimportant, i.e. that the same pdf set will be used for both beams: using separate ones per beam would necessitate a custom \codett{event\_bool\_fn} (or at least a custom \codett{event\_comp\_fn}), as \rex{} does not have support for ordering partons explicitly based on momenta.}\label{code:tearex_pdgrwgt}
\end{algorithm*}

As mentioned above, \tearex{} is a relatively small library when compared to \rex{}, and we hope its usage to be clear from the descriptions above. For a slightly more detailed illustration, see \cref{code:get_comp,code:rwgt_driver} which detail the implementation in \madtrex{}. However, as an illustrative example let us outline an implementation of pdf reweighting using \tearex{}; an example of such a program is shown in \cref{code:tearex_pdgrwgt}, and we will now continue to go through some details of an implementation of this.

The first thing to consider is the subprocess definition; for this illustrative example, we consider only initial-state partons (i.e. particles with \texttt{status=-1}) of which we assume there will always be two, and we further assume that these will be either massless quarks (defined as quarks belonging to the first two generations) or gluons. The three resulting subprocesses are ones with either two quarks, two gluons, or one of each\footnote{The implicit fourth subprocess consisting of any events in our sample that fail these conditions is assumed to be a mismatch for the reweighting procedure and thus is given zero weights for each weight appended to the sample. In a procedure like pdf reweighting this may not be the intention, and one could add an explicit ``all-encompassing'' fourth subprocess using the pre-defined \evbelongs{} returned from \codett{REX::all\_events\_belong()} which, as the name suggests, just returns \codett{true} for all \event{}s. Combining this with a trivial \codett{REX::tea::weightor} that only returns e.g. \codett{std::vector<double> ones(process.events.size(), 1.0)} and placing it as the very last \procrwgt{} ensures any events that have at least one different initial-state parton will maintain their original model weight. The procedure to treat events with only one initial-state quark would necessitate either an \evbelongs{} object with all possible additional initial-state partons \textit{or} a custom \codett{event\_bool\_fn}, neither of which would be particularly difficult to implement.}. As we define our \procrwgt{}s and consequently the \codett{event\_hash\_fn} used to sort the \lhe{} object using the \evbelongs{} type, the transposed \process{} objects will be filtered to only these initial-state partons which will furthermore always have the ordering specified by the \event{}s used in \cref{code:tearex_pdgrwgt}.

\begin{algorithm}
\begin{lstlisting}[language=rex]
class my_pdfs{
private:
   std::vector<REX::tea::weightor> 
     qq_pdfs, gq_pdfs, gg_pdfs;
   size_t curr_pdf;
   bool increment()
   { ++curr_pdf; return true; }
public:
   std::shared_ptr<std::vector<double>>
    qq_pdf(REX::process& p)
    { return qq_pdfs[curr_pdf](p); }
// And similarly for gq_pdf(s), gg_pdf(s)
...
   std::vector<REX::tea::iterator>
    get_iterators()
    { 
      std::vector<REX::tea::iterator>
       iters(qq_pdfs.size(), &increment);
      return iters;
    }
}
\end{lstlisting}
\caption{Minimal illustration of a pdf reweighting helper class to modify pdf sets globally rather than supplying a vector of function pointers to individual pdf sets.}\label{code:my_pdfs_helper}
\end{algorithm}

Once the \evbelongs{} objects are defined, corresponding \codett{REX::tea::weightor}s need to be loaded. The example in \cref{code:tearex_pdgrwgt} only supplies a single \weightor{}, but these could equally well be \codett{std::vector}s of \weightor{}s with entries for each pdf set. In that case, the \procrwgt{} member \codett{normaliser} should also be set to define the original weight with which reweighting is performed with respect to. Alternatively, the format shown in \cref{code:tearex_pdgrwgt} can be used alongside a global wrapper and iterator which is cycled through using the \rwgtr{} member \iterator{}s, as shown in \cref{code:my_pdfs_helper}, where in this minimal implementation it is clear that precautions need to be taken with regard to the sizes of the vectors of pdf sets, and we note that \rwgtr{} calls \iterator{}s \textit{before} running \weightor{}s, meaning in \cref{code:my_pdfs_helper} the first element of the vectors of functions should be the original pdf set. Implementations using global function wrappers and \iterator{}s are likely to be slower than ones with vectors of \weightor{}s due to the required sync between \rwgtr{} and \procrwgt{}s, but may be simpler or necessary depending on the specific structure of the reweighting functions. For pdf reweighting, specifically, this should not be an issue, but for e.g. \madtrex{} where scattering amplitude routines read physics parameters from disk keeping iterations in sync is imperative.

Once the \procrwgt{}s have been defined and the \lhe{} object initialised, the \rwgtr{} can be constructed directly using the explicit \rwgtr{} constructors, and all the intricacies of sorting and transposing events, running reweighting iterations and normalising, and appending the new weights to the \lhe{} are done automatically by calling the \codett{reweightor::run()} function.

\section{\mg{} \tearex{} reweighting executables}
\label{sec:madtrex}

\madtrex{} is an extension to the \cudacpp{} plugin \cite{Valassi:2021ljk,Valassi:2022dkc,Valassi:2023yud,Hageboeck:2023blb,Valassi:2025xfn,Hagebock:2025jyk} for \madgraph{} (\mg{}) \cite{Alwall:2014hca} repurposing the scattering amplitude routines written for data-parallel event generation as a basis for data-parallel event reweighting using the \tearex{} library. Specifically, \madtrex{} enables model parameter reweighting with an alternate backend for the \mg{} reweighting module \cite{Mattelaer:2016gcx} built with \tearex{} --- specifically the SLHA backend presented in \cref{sec:tearex_slha} --- and compiled libraries of the process-specific scattering amplitudes generated by \cudacpp{}.

Below, we detail the usage of and speed-up provided by \madtrex{} when compared to \mg{} reweighting. \Cref{sec:madtrex_manual} provides an in-depth manual for using \madtrex{} in the context of the \cudacpp{} plugin, including installation, usage, and for the interested reader a description of the underlying implementation. In \cref{sec:madtrex_results} we then present runtime comparisons between \madtrex{} and the default \mg{} module, including some discussion on the different sources of speed-up of which there are several beyond the hardware acceleration provided by \cudacpp{} scattering amplitudes.

\subsection{Manual}
\label{sec:madtrex_manual}

While \madtrex{} uses the exact same interface as (generic) reweighting in \mg{}, there are some details worth mentioning. This manual describes the installation of \madtrex{} in \cref{sec:madtrex_install} and how to use it for parameter reweighting in \cref{sec:madtrex_usage}. For details on the implementation of the reweighting executable program, see \cref{sec:madtrex_implementation}.

\subsubsection{Installation}
\label{sec:madtrex_install}

\madtrex{} has been integrated into the \cudacpp{} main repository alongside copies of the 1.0.0 releases of \rex{} and \tearex{}, and will be included in all upcoming releases corresponding to \mg{} v3.6.4 and onward. For more extensive details on installing \cudacpp{} refer to \cite{Hagebock:2025jyk}, but we note that as of \mg{} version 3.6.0 \cudacpp{} can be installed directly through the \mg{} CLI using the command
\begin{lstlisting}[language=bash]
MG5_aMC> install cudacpp 
\end{lstlisting}
with the optional additional argument \texttt{--cudacpp\_tarball=URL} with \texttt{URL} the URL of a specific \cudacpp{} release provided as a tarball.

For the time being, updates to \rex{}/\tearex{} are not automatically propagated to the \cudacpp{} repository; furthermore, manual updates of the copies provided with \cudacpp{} are only anticipated for major \rex{} or \tearex{} releases and even then likely only if the updates are expected to improve \madtrex{} performance explicitly. However, alternate versions of these libraries can of course be manually installed by the end-user.

Changing \rex{} and \tearex{} releases is as simple as overwriting the existing ones and recompiling. The files in question are \texttt{Rex.h}, \texttt{Rex.cc}, \texttt{teaRex.h}, and \texttt{teaRex.cc}, all stored in the directory
\begin{lstlisting}[language=bash]
/mg5amcnlo/PLUGIN/CUDACPP_OUTPUT/MadtRex/
\end{lstlisting}
and can be compiled using the minimal command \texttt{make -f rex.mk}.

\subsubsection{Usage}
\label{sec:madtrex_usage}

\madtrex{} uses the same interface as \mg{} reweighting, although it has some restrictions that the latter lacks. Unlike \mg{}, which supports reweighting at both leading and next-to-leading order, \madtrex{} is restricted to leading order reweighting; furthermore, the default reweighting mode in \mg{} is helicity-exclusive (i.e. the reweighted event is only evaluated at the same helicity configuration as in the original model), whereas \madtrex{} is limited to helicity-summed reweighting in both the original and reweighted model due to the lack of helicity-specific scattering amplitudes supported by \cudacpp{}. Aside from these restrictions, \madtrex{} supports reweighting to and from any leading order (tree-level) model supported by the \cudacpp{} plugin.

Once installed, \madtrex{} reweighting can be enabled by setting the \mg{} \texttt{reweight} flag to \texttt{madtrex} at program launch, i.e. once in the \mg{} command line interface running the commands
\begin{lstlisting}
generate PROCESS
output DIRECTORY
launch
reweight=madtrex
...
\end{lstlisting}
where the \texttt{reweight\_card.dat} can then be modified to include values for the desired hardware backend, floating point precision, and number of CPU threads to launch from the host executable, using the commands
\begin{lstlisting}
change backend BACKEND
change fptype FPTYPE
change nb_thread NB_THREAD
\end{lstlisting}
which must be appended before the first \texttt{launch} command. Any strictly positive integer is allowed for \texttt{nb\_thread}, while supported options for the compile-time arguments are
\begin{itemize}
    \item \texttt{backend}: \texttt{cppauto, cppnone, cppsse4, cppavx2, cpp512y, cpp512z, cuda, hip}. To reweight on a SIMT GPU, set the backend to the corresponding framework (i.e. CUDA for Nvidia GPUs or HIP for AMD GPUs), otherwise we recommend using the default \texttt{cppauto} which automatically detects the best SIMD instructions supported by the machine.
    \item  \texttt{fptype}: \texttt{m} (mixed precision), \texttt{d} (FP64), and \texttt{f} (FP32). Default is \texttt{m}, which computes scattering amplitudes in FP64 and colour algebra in FP32. We recommend avoiding \texttt{f} due to the risk of catastrophic cancellations between Feynman diagrams, but leave the option available.
\end{itemize}
For further details on these options, consult \cudacpp{} documentation \cite{Hagebock:2025jyk}.

Once the reweighting card has been set, keep running the event generation as normal and \madtrex{} will take care of the rest. If no compiled versions of \rex{} or \tearex{} are detected, they will be compiled before code generation for the \madtrex{} executable. This may take some time but only needs to be done once, after which the library can be linked against for all future \madtrex{} calls on the same machine.

\subsubsection{Backend details}
\label{sec:madtrex_implementation}

\begin{algorithm*} 
\begin{lstlisting}[language=rex]
std::shared_ptr<REX::eventBelongs> get_comp()
{ static std::vector<std::vector<short int>> stats = {{-1,-1,1,1},{-1,-1,1,1}};
  static std::vector<std::vector<long int>> pdgs = {{-11,11,-11,11},{-13,13,-13,13}};
  static std::vector<std::shared_ptr<REX::event>> loc_evs;
  for (size_t i = 0; i < stats.size(); ++i)
  {  auto ev = std::make_shared<REX::event>(pdgs[i].size());
     ev->set_status(stats[i]);
     ev->set_pdg(pdgs[i]);
     loc_evs.push_back(ev); }
  return std::make_shared<REX::eventBelongs>(loc_evs, REX::external_legs_comparator); }
\end{lstlisting}
\caption{The \codett{get\_comp()} function provided in one of the \texttt{rwgt\_runner.cc} files for \madtrex{} reweighting of the standard model LO process $l^+ l^- \to l^+ l^-$. Note that both the \texttt{stats} and \texttt{pdgs} vectors consist of two elements, corresponding to the two sets of external legs this particular subprocess evaluates, in this case when the initial- and final-state particles are identical.} \label{code:get_comp}
\end{algorithm*}

\madtrex{} executables are relatively simple programs, in the sense that all functionality is provided by \rex{}, \tearex{}, and \cudacpp{}-generated scattering amplitude functions. The reweighting iterations are provided by the \codett{REX::tea::param\_rwgt} type detailed in \cref{sec:tearex_slha}, and we will forego repeating the details here, but do note that the Python CLI driver will translate the \texttt{reweight\_card.dat} to the more specific standard required by \tearex{}, meaning \madtrex{} has support for user-friendly \mg{} commands such as parameter names (e.g. \texttt{aEW} for $\alpha_{EW}$) or scans (i.e. automatic reweighting over several parameter values, e.g. \texttt{set aEW scan:[100,150,200]}).

With these surrounding details already provided, we turn to the implementation of \cudacpp{} scattering amplitudes as distinct libraries to be included within a single \madtrex{} executable. For details on \cudacpp{} itself and the details of how scattering amplitudes are interfaced with the \textsc{MadEvent} event generator, see \cite{Hagebock:2025jyk}. The only necessary points to mention here are that \cudacpp{} generates data-parallel scattering amplitude evaluation routines with a minimal bridge API to allow other programs to access these routines by providing the relevant process data in a column-major SoA format.

Fundamentally, \madtrex{} executables have three parts stored across three separate files: the generic \texttt{rwgt\_instance}, which defines the functionality connecting the \cudacpp{} API and the executable itself; the process-specific \texttt{rwgt\_runner}, providing an \codett{event\_bool\_fn} to specify \procrwgt{}s and a wrapper for the specific scattering amplitude related to that \process{}; and the executable \texttt{rwgt\_driver}, sorting out the different \texttt{rwgt\_runner}s and constructing a \rwgtr{} from them and the LHE file to be reweighted. The latter two are generated by \madtrex{} for a given process to be reweighted --- \texttt{rwgt\_driver} less so than \texttt{rwgt\_runner} --- while the first one simply defines the interface between the latter two. Since the structure of the \texttt{rwgt\_instance} files is primarily to handle the details of the \cudacpp{} API, we forego details.

\begin{algorithm*}
\begin{lstlisting}[language=rexaster]
int main(int argc, char **argv){
   std::cout << "Starting MadtRex driver...\n"
   ...
   static std::vector<std::shared_ptr<REX::tea::procReweightor>> rwgtRun =
      {P1_Sigma_sm_epem_epem::make_reweightor(), 
       P1_Sigma_sm_epem_mupmum::make_reweightor(), 
       P1_Sigma_sm_epmum_epmum::make_reweightor()};
   auto rwgt_runner = REX::tea::param_rwgt(REX::load_lhef(lheFilePath), rwgtRun);
   rwgt_runner.read_slha_rwgt(slhaPath, rwgtCardPath);
   rwgt_runner.pool_threads = nb_threads;
   rwgt_runner.run();
   std::cout << "\nReweighting procedure finished.\n";
   std::ofstream lhe_out(outputPath);
   if (!lhe_out)
      throw std::runtime_error("Failed to open output LHE file for writing.");
   rwgt_runner.print(lhe_out, true);
   std::cout << "Reweighted LHE file written to " << outputPath << ".\n";
   ...
   return 0;
}
\end{lstlisting}
\caption{The \texttt{main} function for the \madtrex{} executable \texttt{rwgt\_driver.cc}, argument handling omitted. All functionality shown is either directly from \rex{}, \tearex{}, or the scattering amplitude wrappers in the \texttt{rwgt\_runner}s.}\label{code:rwgt_driver}
\end{algorithm*}

Turning first to the \texttt{rwgt\_runner.cc} file, it holds exactly four functions, with two of them being functionality wrappers. These are identically named across \texttt{rwgt\_runner}s, but each one is wrapped in a namespace defining its particular subprocess. The first function, \codett{get\_comp}, creates an \evbelongs{} object corresponding to the specific scattering amplitude code for this particular subprocess. Since this will depend entirely on which amplitudes the given subprocess is to evaluate, this function is generated independently for each generated subprocess. An example of this function for reweighting the process $l^+ l^- \to l^+ l^-$ is provided in \cref{code:get_comp}, where an \evbelongs{} object testing whether a given \event{} corresponds to either of the hard scattering processes  $e^+ e^- \to e^+ e^-$ or $\mu^+ \mu^- \to \mu^+ \mu^-$ is constructed.

The next function in \texttt{rwgt\_runner.cc} is \codett{amp}, which is a call to the \cudacpp{} API wrapped in the process-specific namespace to avoid symbol overlap between subprocesses; specifically for parameter reweighting, the only \process{} information passed on to the amplitude routine are particle momenta, accessed through the call \codett{process::pUP().flat_vector()}. The final two functions are both wrappers --- one, \codett{bridgeConstr}, for creating a \texttt{rwgt\_instance} object handling the intricacies of going from a \process{} to the expected arguments for \cudacpp{}; and the other, \codett{make\_reweightor}, creating  a \procrwgt{} from the bridge just mentioned as well as the \codett{get\_comp()} function illustrated in \cref{code:get_comp}. The only one of these functions called directly from \texttt{rwgt\_driver} is \codett{make\_reweightor} --- all other intricacies are handled by \tearex{}\footnote{Do note, however, that the \texttt{rwgt\_runner.cc} files and all the surrounding scattering amplitude functionality it accesses have significant symbol overlap, as \cudacpp{} uses the same names across subprocesses and internally uses no process-specific namespaces. In \madtrex{} this is overcome using the linker flag \texttt{-Bsymbolic}, which binds references to global symbols to the definition within the shared library.}.

Aside from argument and error handling, \texttt{rwgt\_driver.cc} is also a very simple file. At code generation, the corresponding \texttt{rwgt\_runner.h} files are added to the list of included header files, and one additional line is written constructing a vector of \procrwgt{}s consisting of the return value from each \codett{make\_reweightor} function. Returning to the process $l^+ l^- \to l^+ l^-$ and omitting argument handling, the full \madtrex{} executable \codett{main} function is shown in \cref{code:rwgt_driver}. All the executable needs to do is define the \procrwgt{}s, given by an \evbelongs{} object and a \weightor{} (since parameter reweighting using \madtrex{} assumes identical \weightor{}s for normalisation and reweighting); define the \iterator{}s --- in this case, a set of functions overwriting the SLHA parameter card read by the \weightor{}s --- and call \codett{reweightor::run()}. All details are sorted out by \tearex{}, illustrating the ease of use the library provides.

\subsection{Runtime comparisons}
\label{sec:madtrex_results}

As a simple illustration of the speed-up provided by \madtrex{} --- with respect to the various data-parallel backends enabled by \cudacpp{}-provided scattering amplitudes as well as in comparison to the reweighting module provided by \mg{} --- we consider a realistic use case for tree-level parameter reweighting, in reweighting SM samples (generated at arbitrary order, although we here stick to leading order for simplicity) to BSM models, allowing for the study of BSM effects on observables by reweighting simulated samples to new models.

For this benchmark, we turn to SM effective field theory (SMEFT), where the SM is interpreted as an effective field theory of a higher-dimensional model and allows for generic parametrisation of BSM effects, assuming the higher-dimensional model abides by the same global symmetries as the SM \cite{Isidori:2023pyp}. Further details on the SMEFT are unimportant here; it is sufficient to note that the SMEFT includes a plenitude of free parameters but reduces to the SM at lower energies, making it an ideal target for simulation recycling using parameter reweighting.

We will consider 4-top production in the SMEFT, i.e. the process
\begin{align}
    p \, p \to t \, \overline{t} \,  t \, \overline{t} \, + n \; \text{jets},
\end{align}
for $n\in\{0,1\}$ and any massless QCD jets. Furthermore, we will consider both the generic process with $p,j$ any massless QCD parton (``multi-channel'') and the case where $p,j$ are both set to be exclusively gluons (``single-channel''). For these four processes, we first generate SM samples (at leading order for practicality) with between 10 and $10^7$ events. Then, we reweight these samples to various different sets of Wilson coefficients for the top-related couplings in the ``$U(3)_l\times U(3)_e$-symmetric'' SMEFT, for which SMEFTsim provides the UFO model \texttt{SMEFTsim\_topU3l} \cite{Brivio:2017btx}.

\begin{figure*}
    \centering
    \includegraphics[width=0.95\linewidth]{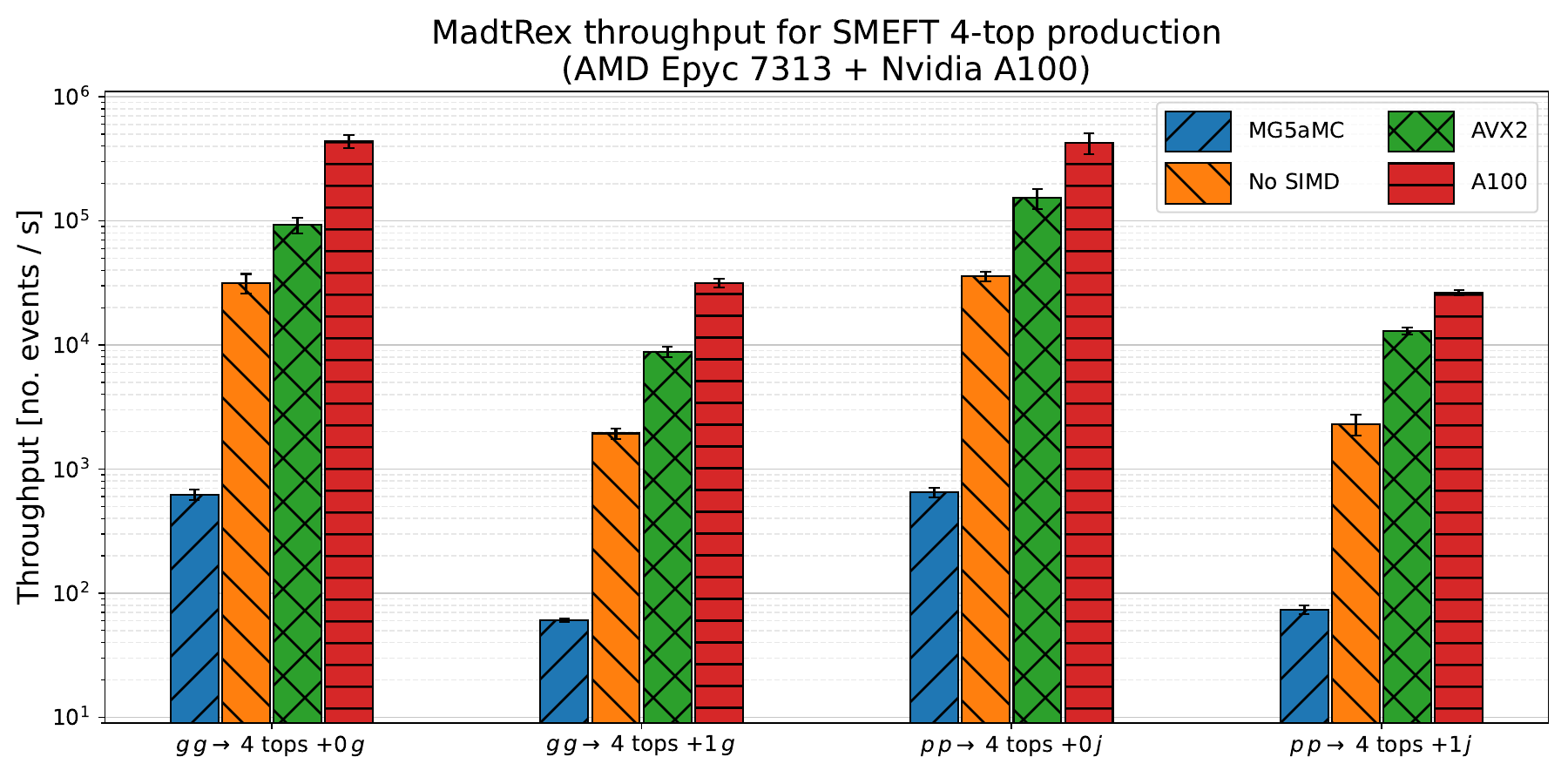}
    \caption{Event throughput for \madtrex{} reweighting as well as the default \mg{} reweighting module for comparison. Throughputs and standard deviations have been calculated based on mean runtimes for various event samples (ranging from 10 to $10^7$ events) with various number of reweighted parameter sets (ranging from 8 to 6435 iterations). Although GPU offloading has a clear advantage over on-host SIMD parallelism, which in turn is faster than scalar instructions, \madtrex{} even without any explicit data parallelism is consistently $\sim40$ times faster than \mg{} reweighting.}
    \label{fig:madtrex_tp}
\end{figure*}

To determine the throughputs of the different implementations --- \mg{} reweighting and \madtrex{} with scalar instructions, SIMD instructions, and GPU offloading --- we start at small sample sizes and few parameter sets reweighted to, and then increase both until a plateau is reached (considered the point at which the throughput is no longer consistently increasing with the number of reweighted events and reweighting iterations). This will of course vary for the different implementations: for \mg{} the plateau was generally reached already for 100 events and 8 reweighting iterations, while for \madtrex{} it was typically necessary to run at least 36 iterations for 1 000, 10 000, and 100 000 events needed to reach the plateau for scalar instructions, AVX2 instructions, and GPU offloading using an Nvidia A100, respectively\footnote{Specifically, reweighted parameter sets were defined in terms of linear to octic power combinations of the considered Wilson coefficients. For \mg{}, no power beyond cubic could be finished within a reasonable time frame, while for \madtrex{} with GPU offloading the octic power combinations would finish for samples of 100 000 events within a couple of hours.}. These measurements are shown in \cref{fig:madtrex_tp}, where \texttt{nb\_thread=1} for all \madtrex{} executions.

\Cref{fig:madtrex_tp} provides extensive insights, but none particularly surprising. Since parameter reweighting is a computationally bound problem completely dominated by scattering amplitude evaluations (the exact part parallelised in \cudacpp{}-generated code) AVX2 instructions provide a speed-up of roughly a factor $4$  compared to scalar instructions. Additionally, for complicated processes like these, GPU offloading using a high-performance general-purpose GPU like the Nvidia A100 can provide further speed-up when compared to on-host SIMD parallelism.

More noteworthy is the sizeable speed-up when comparing scalar \madtrex{} with the native \mg{} reweighting module: For all tested processes, \madtrex{} execution \textit{without} any explicitly implemented data parallelism has a throughput $30-60$ times greater than \mg{}. There are two reasons for this: (1) rather than sorting events online for each reweighting iteration (as \mg{} does), \madtrex{} has a ``one-and-done'' upfront sorting algorithm; and (2) \madtrex{} runs a compiled reweighting executable rather than a Python driver calling Fortran functions through \texttt{f2py}.

Point (1) reduces the leading constant in the linear runtime growth by limiting it to only the number of events, i.e. the sorting runtimes grow as
\begin{align}
    t_{\text{\mg{}}} &= \mathcal{O}\left( \#\text{events} \times \#\text{iterations}  \right),\\
    t_{\text{\madtrex{}}} &= \mathcal{O}(\# \text{events} ),
\end{align}
minimising the runtime cost of reweighting to additional parameter sets.

When considering available computational power scaling, however, point (2) is more interesting: the structure of \mg{} reweighting limits it to running sequentially single-threaded, and overcoming this would require explicit modifications to the reweighting module. With \madtrex{}, however, this is automatically provided through compiler optimisation. As mentioned, the tests shown in \cref{fig:madtrex_tp} were run with \texttt{nb\_thread=1}, meaning no multithreading over subprocesses; however, compiler optimisation can still enable multithreading \textit{within} a given subprocess. This has been directly observed: Running on-CPU \madtrex{} executables, child processes are consistently launched across 8 CPU cores without any multithreading across subprocesses\footnote{Due to how multithreading is set up in \madtrex{}, the single-channel processes with only external gluons aside from the four top quarks cannot benefit from subprocess multithreading since these processes consist of a single subprocess in the \madtrex{} scheme.}. We can assume that a factor $\sim8$ of on-CPU \madtrex{} speed-up thus comes from in-subprocess multithreading provided directly from compiler optimisation, leaving the speed-up from not using an interpreted language at $\sim5$. Whether this can be scaled further will be considered for continued \tearex{} and \madtrex{} development.

However, this benefit does not apply for GPU offloading, as in-subprocess multithreading is already the target of the CUDA-compiled \madtrex{} executables. As evidence of this, consider the following: the single-channel gluonic processes are the most computationally heavy subprocess of the multi-channel processes. Consequently, for a computationally bound problem, the throughput for multi-channel processes should be equal to or greater than the single-channel ones, which is seen for all on-CPU reweighting implementations. On the other hand, the addition of multiple subprocesses, which are run sequentially, will impact a latency-dominated executable, and as \cref{fig:madtrex_tp} shows, the GPU execution for multi-channel processes is consistently slower than single-channel processes. With this in mind, for few-CPU \madtrex{} jobs there is little reason to try to optimise the \texttt{nb\_thread} variable for on-CPU jobs, while it could be essential for making the best use of GPU offloading.

Regardless of the specifics, it is clear that \madtrex{} provides both immediate speed-up when compared to \mg{} reweighting, and great potential for better scalability in across larger and distributed systems. Although \madtrex{} currently only provides functionality for tree-level LO parameter reweighting, this alone enables the reuse of already simulated SM samples for the study of BSM models such as the SMEFT used here. In the long term, further developments in \cudacpp{} functionality towards NLO event generation could enable NLO parameter reweighting in \madtrex{} with minimal development necessary from the \madtrex{} side.

\section{Conclusions}
\label{sec:conclusions}

The three codes presented in this paper --- \rex{}, \tearex{}, and \madtrex{} --- provide an accessible entry point for HEP software handling parton-level hard scattering events. \rex{} provides a physics-oriented interface for LHE file format events while providing tools for the simple implementation of I/O for further LHE-like file formats, and furthermore enables trivial transposition between human-readable OO data formats and SoA formats designed for data-parallel hardware acceleration. These data formats are used as a basis for completely generic event reweighting in \tearex{}, which provides a basis structure for reweighting events to arbitrary conditions, necessitating users to only provide the corresponding reweighting functions. Using \rex{} for event data handling and \tearex{} as a basis, the \madtrex{} reweighting module enables data-parallel model parameter reweighting within \mg{} using the \cudacpp{} plugin as a basis for scattering amplitude evaluations, and only using the \rex{} sorting algorithm and compiler optimisation consistently achieves reweighting throughputs $30-60$ times greater than the default \mg{} reweighting module for computationally complex processes; using on-CPU SIMD instructions increases this throughput further by the expected maximal gain for AVX2 instructions, and GPU offloading can push the total acceleration up to a factor $300-700$ depending on the process. \rex{} and \tearex{} are currently available on GitHub at the URL \href{https://github.com/zeniheisser/Rex}{https://github.com/zeniheisser/Rex}, and as of the next \cudacpp{} version \madtrex{} will be included as part of the \cudacpp{} plugin alongside versions 1.0.0 of \rex{} and \tearex{}.

Going forward, all three codes have great potential for further development. Starting with \rex{}, ensuring the data access interface is as simple as possible is and will always be the main concern, although what exactly this entails remains to be seen based on user feedback. Some possibilities, though, include bindings for more commonly used programming languages --- particularly Python --- allowing its usage across a far wider range of software than just compiled C++ programs. Additional data access functionality is also simple to implement and will be considered upon request. One particular point of further consideration is whether to attempt to optimise \rex{} for memory consumption; at present, \rex{} data takes up roughly the same size in memory as the LHE plaintext format does on disk, which may or may not be a limiting factor depending on the sizes of samples used for practical applications.

While minimal in size, \tearex{} has already proven to be extremely potent at its intended purpose of enabling simple implementation of (data-parallel) parton-level event reweighting. Furthermore, being an extension to \rex{}, \tearex{} will benefit directly from any additional development in \rex{}. Considering specifically \tearex{} though, there are some potential avenues of further development: first, \tearex{} only provides explicit multithreading support across separately defined subprocesses of a given event sample; this does not necessarily make the best use of available compute, and applying further subprocess splitting where subprocesses with many events are divided into additional separate execution tracks may make better use of on-CPU multithreading. While we have not identified further optimisations in the \tearex{} structure itself, we are open to user suggestions; however, we expect ease of use to be a more interesting concern, just as for \rex{}. Like we suggested for \rex{}, we expect Python bindings (or similar) to be a valuable future development to allow for the automatic data-parallel reweighting of generic reweighting using \textit{any} input functions without needing to implement an executable program.

Further \madtrex{} development is unlikely to focus on optimisation; \cudacpp{} scattering amplitudes are already perfectly parallel and by virtue of only calling the scattering amplitudes themselves \madtrex{} makes perfect use of them. Of course, \madtrex{} will benefit from any optimisation in \cudacpp{} amplitudes, but more importantly it can gain extended functionality from added features to \cudacpp{}. Of greater interest for the average user, though, may be that \cudacpp{} is planned to become the default (leading order) code generator for \textsc{MadGraph} as part of the upcoming \textsc{MadGraph7} project, and that there is active discussion about making \madtrex{} the default reweighting module as part of that development. Thus, further \madtrex{} development is likely to be focused on integrating it further into the \mg{} architecture and extending its functionality to treat other reweighting use cases within the \mg{} suite, such as NLO parameter reweighting or pdf reweighting.

Overall, \rex{} and \tearex{} provide an efficient parton-level event interface for HEP software, enabling the trivial transposition between OO and SoA data formats as well as the generic reweighting of events using completely generic functions for whatever parameters are being reweighted. Implementing this as well as proving its applicability, the \madtrex{} reweighting module for \mg{} can provide a $30-50$ times throughput increase for computationally heavy processes on the exact same CPU without any explicitly implemented data parallelism, while AVX2 instructions increase this to a factor $150-300$ speed-up and GPU offloading using an Nvidia A100 GPU can push it as far as a factor $700$ throughput increase. Further developments in all codes are likely to primarily consider functionality and ease of access as well as potential integration into existing codebases in order to provide these benefits for as large a fraction of the community as possible.

\section*{Acknowledgements}

We extend our gratitude to Olivier Mattelaer for discussions regarding the interfacing used in \rex{} and \tearex{}, and in particular for his assistance in enabling integration between \madgraph{} and \madtrex{}; additional thanks are extended to Andrea Valassi for his assistance in the interfacing between \tearex{}, \madtrex{}, and the \cudacpp{} plugin; and to Stephan Hageb{\"o}ck for discussion and recommendations regarding efficient data interfacing between different data formats with respect to LHE parsing and storage within \rex{}. Additionally, we thank all contributors whose work has directly and indirectly impacted \cudacpp{} development, as well as all \mg{} authors, past and present. Computational resources were partially provided by the Calcul Intensif et Stockage de Masse (CISM) technological platform. SR and ZW acknowledge support from CERN openlab as well as the Next Generation Triggers project hosted by CERN, which is funded by the Eric and Wendy Schmidt Fund for Strategic Innovation.

\printbibliography


\end{document}